\documentclass[]{aastex61}
\usepackage{amsmath}
\def\new#1{{#1}}
\def\newest#1{{#1}}

\accepted{June 12, 2019}
\submitjournal{\it The Astrophysical Journal}

\shorttitle{Habitability of S-type tidal-locked planets}
\shortauthors{Okuya et al.}

\begin{document}

\title{Effects of a binary companion star on habitability of tidally locked planets around an M-type host star}

\author[0000-0002-0786-7307]{Ayaka Okuya}
\affil{Department of Earth and Planetary Sciences, Tokyo Institute of Technology, Ookayama, Meguro, Tokyo 152-8551, Japan}

\author{Yuka Fujii}
\affiliation{Earth-Life Science Institute, Tokyo Institute of Technology, Ookayama, Meguro, Tokyo 152-8550, Japan}


\author{Shigeru Ida}
\affiliation{Earth-Life Science Institute, Tokyo Institute of Technology, Ookayama, Meguro, Tokyo 152-8550, Japan}



\begin{abstract}
Planets in the ``Habitable Zones'' around M-type stars are important
targets for characterization in future observations.
Due to tidal-locking in synchronous spin-orbit rotations, 
the planets tend to have a hot dayside and a cold nightside. 
On the cold nightside, water vapor transferred from the dayside can be frozen in (``cold trap'') or the major atmospheric constituent could also condense (``atmospheric collapse'') if the atmosphere is so thin that the heat re-distribution is not efficient, in the case of a single M-type star. 
Motivated by the abundance of binary star systems, 
we investigate the effects of irradiation from a G-type companion star
on the climate of a tidally locked planet around an M-type star 
using the 2D energy balance model.
We find that the irradiation from the G-type star is more effective at warming up the nightside of the planet than the dayside.
This contributes to the prevention of the irreversible trapping of water and atmosphere on the cold nightside, broadening the parameter space where tidally locked planets can maintain surface liquid water.
Tidally locked ocean planets with $\lesssim$ 0.3 bar atmospheres or land planets with $\lesssim 3$ bar atmospheres can realize temperate climate with surface liquid water only when they are also irradiated by a companion star with a separation of 1 - 4 au.
We also demonstrate that planets with given properties can be in the Earth-like temperate climate regime or in a completely frozen state under the same total irradiation.

\end{abstract}



\section{Introduction} \label{sec:intro}
Planets in the \new{so-called ``Habitable Zones''} (HZs), where liquid water can exist on the planetary surface,
around M-type stars are easier to detect through radial velocity surveys
owing to the smaller stellar mass and HZs closer to the star.
Their small stellar size also has an advantage in transit detection of small planets.
The Earth-sized planets recently discovered around the HZs, TRAPPIST-1 e, f, and g, Proxima centauri b, and LHS 1140b, 
orbit M-type stars. 
Future observations with the James Webb Space Telescope and ground-based extremely large telescopes will aim to characterize the atmospheres of these planets around M-type stars to search for habitable conditions and eventually for biosignatures.

These planets are likely to be tidally locked due to their proximity to the host stars \citep{Kasting+1993}, and to have a fixed warm/hot dayside and cold nightside. 
Non-sunlight in the cold hemisphere poses at least two potential problems for the habitability of the planet:
i) ``atmospheric collapse" and ii) ``cold trap" of surface water. 
If the local temperature on the nightside is so low that the major atmospheric constituent condenses out, the loss of the greenhouse effect and heat transport would cause further cooling, and the planet would undergo a transition into a cold state with a thin atmosphere. This phenomenon is called ``atmospheric collapse" and 
has been considered an obstacle to habitability \citep[e.g.,][]{Joshi+1997}. 
In addition, on the planets with a limited small amount of surface water (``land planets"), 
the water is transported from warmer regions to cooler ones by atmospheric circulation \citep{Abe+2005, Abe+2011}. 
In the tidally locked land planet, the dayside would be left free from water, and all of the water would be frozen on the nightside \citep{Leconte+2013}. 
The ``cold trap" of water would be irreversible unless ice flow driven by gravity or internal thermal flux is strong enough \citep{Leconte+2013, Turbet+2016, Turbet+2017}.

If the planet-hosting M-type star has a much brighter stellar companion such as a G-type star, it periodically irradiates the cold nightside of a tidally locked planet around the M-type star. 
Such a configuration may rescue HZ planets from the above-mentioned difficulties if the binary separation is appropriate: close enough for the irradiation of the companion star to affect the planetary climate, but not too close to ensure the stability of the planetary orbit. In reality, systems comprised of an M-type star and G-type star are not rare. 
About half of all G-type stars in the solar neighborhood have binary companions and 
the number distribution of their mass ratio ($q$), $dN/dq$, is approximately constant \citep{Raghavan+2010}.
In other words, a substantial fraction of G-type stars have M-type companion stars.

Circumsteller planets in binary systems like those described above are called ``S-type" planets, as opposed to the circumbinary planets called ``P-type" planets. 
More than 60 S-type exoplanets are
known today. While most of them are wide binaries,
a relatively close binary system such as Kepler 420 A and B, with a separation of 5.3 au has an S-type eccentric giant planet with semimajor axis 0.38 au around Kepler 420 A \citep{Santerne+2014}. 
Although it is not easy to detect S-type planets in close binary systems,
future surveys may reveal the occurrence rate of S-type planets. 
For example, \citet{Oshagh+2017} proposed a new detection method for S-type planets in eclipsing binaries by using a correlation between the stellar radial velocities (RVs), eclipse timing variations (ETVs), and eclipse duration variations (EDVs). 
Whether S-type planets in close binaries are common or not is an active field of research from the viewpoint of planet formation. 
Circumstellar disks can exist if the disk radius is smaller than $\sim 0.2-0.3$
of the binary separation, and gas accretion from the circumbinary disk
to the individual circumstellar disks may exist \citep{Artymowicz+Lubow1994}.
It may be possible that the S-type planets are formed 
in the stable regions of these disks, although many issues remain to be studied
\citep[e.g.,][]{Thebault2015,Dupuy+2016,Gong+2018}.
We leave the formation of S-type planets in relatively close binary systems
for future studies.

Some previous studies \citep{Kaltenegger+2013, Jaime+2014} have considered the habitability of S-type planets, by extending the HZs of single stars obtained with 1D modeling of planetary atmospheres \citep[e.g.][]{Kasting+1993, Kopparapu+2013}.  
Their estimates of HZs of S-type planets are based on the total irradiance the planet receives from both stars and the orbital stability condition, and did not take into account the horizontal \new{dimension} of the planetary surface. 
However, as we pointed out above, investigations into the habitability of planets should take into account the effects of atmospheric collapse and cold trap, and therefore the global structure of planetary surface temperature is essential. 
An approach to address these effects is GCM (General Circulation Model) simulations where individual physical and chemical processes including radiative transfer, atmospheric/oceanic dynamics, and phase transition of water are calculated on the three-dimensional grids; GCM simulations have been applied to tidally locked planets around single M-type stars \citep[e.g.,][]{Turbet+2016, Turbet+2017,Kopparapu+2017,Fujii_et_al_2017}.
An alternative approach is the Energy Balance Model (EBM), which finds the planetary surface temperature distribution by solving simple horizontal energy transfer across the planetary surface. 
While EBM greatly simplifies or ignores the individual physical and chemical processes that control the energy transfer, EBMs have been useful to study basic climatological properties of exoplanets \citep{David+2009,David+2010,Checlair+2017}.

In this paper, we study the effects of irradiation from a G-type companion star on the condition of habitability of tidally locked planets around an M-type star (S-type planets),
taking into account the effects of atmospheric collapse and cold trap.
In order to gain insights into the first-order behavior of the planetary climate exploring a broad parameter space,
we use two-dimensional EBM calculations \citep[e.g.][]{North1975} rather than
complex and computationally expensive GCM simulations.
The planet is assumed to be either fully covered with water (``ocean-covered'') or to have a limited amount of water with most of the surface being bare (``land-covered''), and its atmosphere is either Earth-like or CO$_{2}$-dominated. For each class of planets, we estimate the binary separation that allows for \new{the presence of liquid water on their surfaces}.

In Section \ref{sec:model}, we describe our assumptions on the binary system, the energy balance model used to calculate the the planetary surface temperature distribution,and our \new{criteria for atmospheric collapse and the cold trap} based on the planetary surface temperature distribution.
In Section \ref{sec:results}, we demonstrate the surface temperature maps with and without a G-type companion star, and analyze the behavior of temperature on \new{ocean- or land-covered} planets by changing binary separations. Finally, we present the \new{orbital region where planets of different types can maintain temperate climate}  and compare them to the case of a planet around a single M-type star without a companion star. 
We discuss parameters that would affect our results and observability of the planets we focus on in Section \ref{sec:discussion}, and summarize our findings in Section \ref{sec:conclusion}.

\section{Model} \label{sec:model}

In section \ref{subsec:config}, we explain the settings of the binary stars and the S-type planet that we simulate.
In section \ref{subsec:EBM}, we describe the two-dimensional EBM for the planet and the input parameters.
Section \ref{subsec:howHZ} introduces our criteria for \new{the atmospheric collapse and cold trap.}

\subsection{Assumed System Architecture} \label{subsec:config} 

We consider binary systems composed of a G2V and an M3V main-sequence star whose basic parameters are summarized in Table \ref{tab:stellar_properties}: 
the G-type star has the luminosity $L_{\rm G} = 1L_{\odot}$ and the mass $M_{\rm G}=1M_{\odot}$, and the M-type star has $L_{\rm M} = 0.01L_{\odot}$ and $M_{\rm M}=0.25M_{\odot}$, consistent with the mass-luminosity relation of M-type stars \citep[e.g.,][]{Boyajian+2012}.
We change the binary separation between the G-type star and the M-type star from 0.1 au to 5.5 au by 0.1 au.

In most of the calculations in this paper, 
we assume that the binary eccentricity, $e_{\rm{B}}$, is zero for simplicity.
Observations show that the median eccentricity for binary periods of 10 - 1000 days
is $e_{\rm{B}} \sim 0.3$ \citep{Duquennoy+1991}.
We will discuss the case of $e_{\rm{B}}=0.3$ in Section \ref{subsec:eccentricity}. 
In addition, we set the binary inclination relative to the planetary orbital plane as $i_{\rm B} = 0$ for simplicity. 
The discussion on the effects of non-zero $i_{\rm B}$ is left for future work.

\begin{deluxetable*}{cccc}[h]
\tablecaption{Stellar properties of the binary stars \label{tab:stellar_properties}}
\tablecolumns{5}
\tablenum{1}
\tablewidth{0pt}
\tablehead{
\colhead{Spectral Type} &
\colhead{$T_{\rm eff}$} &
\colhead{Luminosity $L_*$  [$L_{\odot}$]} &
\colhead{Mass $M_*$ [$M_{\odot}$]}
}
\startdata
G2V & 5778 K & 1 & 1\\
M3V & 3300 K & 0.01 & 0.25\\
\enddata
\end{deluxetable*}

We assume that the M-type star is orbited by a rocky planet. The mass and radius of the planet are set at Earth's values. 
The semimajor axis of the planet is changed within the range of $a < a_{\rm max}$ where $a_{\rm max}$ is the maximum semimajor axis for the planetary orbit not to be destabilized by the secular perturbations from the G-type star.
We use the fitting formula by \citet{Pichardo+2005}:
\begin{equation}
a_{\rm max} \simeq 0.6 \, a_{\rm B} \, (1-e_{\rm B})^{1.2} \times 
\frac{(M_{\rm M}/(M_{\rm G}+M_{\rm M}))^{0.07}}{1+1.67 (M_{\rm M}/M_{\rm G})^{-2/3} \ln (1+ (M_{\rm M}/M_{\rm G})^{1/3})} \label{eq:stableorbit}
\end{equation}
where $a_{\rm B}$ is the binary separation, and $M_{\rm{M}}$ and $M_{\rm{G}}$
are the masses of the M-type and G-type stars.
For $M_{\rm M}=0.25 \, M_{\rm G}$, 
\begin{equation}
a_{\rm max} \simeq 0.2 \, a_{\rm B} \, (1-e_{\rm B})^{1.2}. \label{eq:a_max}
\end{equation}

We assume a circular planetary orbit ($e=0$) and zero obliquity because of the tidal dissipation in the planet.
If the binary eccentricity is not equal to 0, 
the planetary eccentricity may oscillate.
Within the limits of weak tidal dissipation, the maximum value of the oscillating eccentricity is 
\citep[e.g.,][]{Murray1998},
\begin{align}
e_{\rm{max}}\simeq \frac{5}{2} \frac{a}{a_{\rm B}} e_{\rm B}.
\end{align}
For  $a < a_{\rm max}$, $e_{\rm max} \simeq 0.5 (1-e_{\rm B})^{1.2} e_{\rm B}$. 
Even in the case of $e_{\rm B}=0.3$,  $e_{\rm max} \la 0.1$, which may be negligible.

We postulate that a planet is tidally locked in a 1:1 spin-orbit state.
The tidal-locking limit for a M3V star is estimated to be $\sim$ 0.3 au, and \new{we confine our study to this range, consistent with the postulation. As we will see later, the orbital region where planets have temperate climate are} mostly within this limit.

\subsection{Energy Balance Model} \label{subsec:EBM}

We use an Energy Balance Model (EBM) to study a time-dependent temperature distribution of a tidally locked rocky planet orbiting an M-type star and having a G-type companion star.

An EBM has been widely used to study the climate of the Earth \citep[e.g.][]{North1975} and Mars \citep[e.g.][]{James+1982}. 
EBM solves the planetary surface temperature distribution taking into account the local net radiation flux and the horizontal heat transport; detailed processes including the vertical profile of the atmosphere and phase transition of water are not explicitly solved. 
This is in contrast to General Circulation Models (GCMs) where these processes are parameterized and solved at each 3-dimensional (2 for horizontal, 1 for vertical) grid cell. Because of such simplification, the results from EBM may not be quantitatively accurate. However, EBM is useful in revealing the planetary climate's global trend in response to external forces, and EBM calculations are analytically more tractable.
A much broader parameter space can be surveyed and it is easier to reveal
intrinsic physics with EBM, if the model is properly calibrated by the GCM simulations. We will calibrate our EBM calculations with the results of
GCM simulations for tidally locked planets around single M-type stars 
by \citet{Turbet+2016, Turbet+2017}.

In order to take into account not only the static irradiation from the M-type star, but also the periodic irradiation from the G-type companion star,
we adopt a time-dependent two-dimensional (latitude $\theta$ and longitude $\phi$) EBM, based on \citet{North1975}.
We use $4^{\circ} \times 4^{\circ} $ grids. 
The energy balance equation is 
\begin{equation}
C \frac{\partial T(\theta,\phi,t)}{\partial t}=Q(\theta,\phi,t)-I (\theta,\phi,t)+ \nabla \cdot (\kappa\nabla T(\theta,\phi,t)) ,       \label{eq:EBeq}
\end{equation}
where $T$ is the planetary surface temperature, $t$ is time, 
and $C$ is the heat capacity of the surface, \new{$Q$ is heating by the host star and the companion stars, $I$ is thermal outgoing radiation, and $\kappa $ is the diffusion coefficient. 
The heating, $Q$, is a sum of the time-independent incoming irradiation flux from the M-type star, $F_{\rm{M}}(\theta,\phi)$,
and the time-dependent (periodic) one from the G-type stars, $F_{\rm{G}}(\theta,\phi,t)$,
\begin{equation}
Q(\theta,\phi,t)=F_{\rm{M}}(\theta,\phi)\cdot(1-\alpha_{\rm{M}}(\theta,\phi))+F_{\rm{G}}(\theta,\phi,t)\cdot(1-\alpha_{\rm{G}}(\theta,\phi)),
\label{eq:radiationQ}
\end{equation}}
\new{where $F_{\rm{M}}$ and $F_{\rm{G}}$ are the irradiance by the M-type star and the G-type companion star, respectively, and $\alpha_{\rm{M}}$ and $\alpha_{\rm{G}}$ are corresponding albedos. 
With the input parameters described below, Equation (\ref{eq:EBeq}) is solved} under the boundary condition with no heat transport at the poles for $\theta$ and the periodic boundary condition for $\phi$. 
The numerical calculations continue running until an equilibrium periodical cycle is achieved.

The heat capacity ($C$), albedo ($\alpha $), outgoing thermal flux ($I$), and diffusion coefficient $\kappa$ \new{are determined as described below, } depend\new{ing} on the surface and atmospheric conditions. 
In this paper, we consider \new{the} combinations of two surface types and two atmospheric types. 
For the surface environment, we consider two limiting cases: rocky planets wholly covered with ocean, ``ocean planets, and dry planets with a mostly bare surface but with a small amount of water, ``land planets.'' 
For the atmospheric condition, \new{either} an Earth-like atmosphere (composed of N$_2$ and O$_2$ with 376 ppm CO$_{2}$) or CO$_{2}$-dominated 0.3-10 bar atmosphere is assumed. 
\new{In summary, we consider the following four types:}

\begin{itemize}
\item[(\underline{OE})] \new{ocean planets with} Earth-like atmospheres with 1 bar \new{mixture} of N$_{2}$ and O$_2$ with 376 ppm CO$_{2}$ \new{with varying amount of water vapor},
\item[(\underline{OC})] \new{ocean planets with} CO$_{2}$\new{-dominated atmospheres} of $p = 0.3, 1$ and 2 bar, with varying amount of water vapor, 
\item[(\underline{LE})] \new{land planets with} Earth-like dry atmospheres (same as (OE) but without \new{water vapor}), and 
\item[(\underline{LC})] \new{land planets with} pure CO$_{2}$ atmospheres of $p = 0.3, 1, 3$, and 10 bar. 
\end{itemize}

\new{From a point of view of the planetary formation, land planets are potentially important targets in future observations searching for habitable worlds,
especially around M-type stars. 
Unlike G-type stars, M-type stars experience a prolonged pre-main-sequence stage with an order of magnitude higher luminosity than that in their main-sequence stage. 
During this stage, planets that currently reside in the HZ would have been exposed to extreme irradiation and would have lost a significant amount of the water it originally had (if any) \citep{Ramirez+2014,Tian+2015,Luger+2015}. 
Thus, the substantial number of an planets in the HZ of an M-type star may be desert planets \citep{Tian+2015}, and later delivery of a small amount of water will then make them land planets. }

\subsubsection{Heat Capacity} \label{subsubsec:C}

The values for $C$ for ocean and land planets are adopted from the Earth's values for ocean and land, respectively, which are $C_{\rm ocean}=2.09\times10^{8} \, {\rm J m}^{-2}{\rm K}^{-1}$ and $C_{\rm land}=8.37\times10^{6} \, {\rm J m}^{-2}{\rm K}^{-1}$ \citep{Pollard1983}. 
\new{Over sea-ice, $C_{\rm ice}$ takes twice value of $C_{\rm land}$ for $T<273$K.}
The values for other parameters will be discussed in Section \ref{subsubsec:Q}, \ref{subsubsec:I}, and \ref{subsubsec:D} below.

\subsubsection{Irradiation} \label{subsubsec:Q}

\begin{figure}[ht!]
\centering
\includegraphics[bb=0.000000 0.000000 917.000000 457.000000, width=0.85\textwidth]{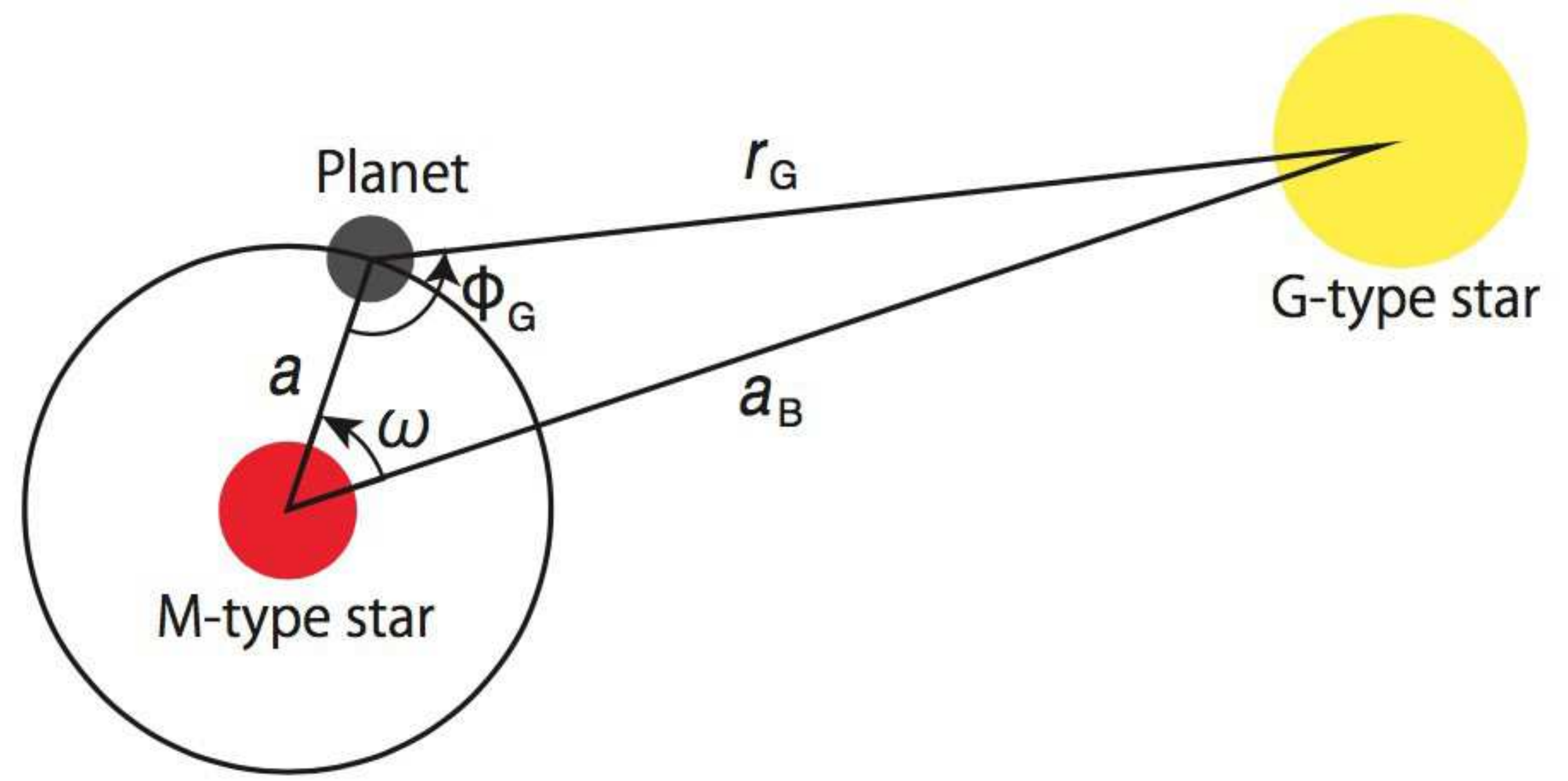}
\caption{
The configuration of the binary stellar system and the planet.
The host star for the planet is the M-type star and the companion star is the G-type star.
The binary separation is $a_{\rm B}$, the planetary semimajor axis is $a$.
The distance $r_{\rm G}$ between the planet and the G-type star and angles $\omega$
and $\phi_{\rm G}$ are given by Eqs.~(\ref{eq:rg}), (\ref{eq:omega}) and (\ref{eq:phig}).
}
\label{fig:3body}
\end{figure}

The irradiance from the two stars on the location $(\theta, \phi)$ of the planetary surface is given by
\begin{eqnarray}
F_{\rm M}(\theta,\phi) &=& \mbox{max}\left\{ \frac{L_{\rm M}}{4\pi a^2} \cos\theta\cos\phi, \;\; 0 \right\} \\
F_{\rm G}(\theta,\phi,t) &=& \mbox{max}\left\{\frac{L_{\rm{G}}}{4\pi \, r_{\rm G}^{2}(t)}\cos\theta\cos(\phi - \phi_{\rm G}(t)) , \;\; 0 \right\}
\end{eqnarray}
\new{where} $r_{\rm G}$ is the distance between the planet and the G-type star
(see Figure \ref{fig:3body}), given by
\begin{equation}
r_{\rm{G}}(t)=\sqrt{a_{\rm B}^{2} + a^{2} - 2 a_{\rm B} \cdot a \cdot \cos{\omega}},
\label{eq:rg}
\end{equation}
$\omega$ is the angle between the direction to the planet and that to the G-type star
from the M-type star,
\begin{equation}
\omega = (\Omega_{\rm p} - \Omega_{\rm G}) t,
\label{eq:omega}
\end{equation}
and $\Omega_{\rm p}$ and $\Omega_{\rm G}$ are
Keplerian frequencies of the planet and G-type star, respectively. 
The longitude of the substellar point of the G-type star is given by
\begin{equation}
\sin{\phi_{\rm G}(t)} = \frac{a_{\rm B}}{r_{\rm G}(t)} \sin{\omega}  \label{eq:phig}.
\end{equation}
As Eq.~(\ref{eq:omega}) shows, 
both $r_G$ and $\phi _G$ oscillate with the synodic period between the planet and the G-type star
relative to the M-type star, causing the periodic change in the insolation pattern of the planet.

\subsubsection{Albedo} \label{subsubsec:alb}
\new{The albedos in Eq.~(\ref{eq:radiationQ}), $\alpha_{\rm{M}}$ and $\alpha_{\rm{G}}$, depend on the planetary surface and atmospheric composition \new{and} pressure.}
The values we used are summarized in Table \ref{tab:experiment} \new{and the assumptions are detailed below}.

The albedo of a cloud-free atmosphere with an underlying surface is generally given by the following combination of \new{albedo} of the atmosphere ($\alpha_{\rm{atm}}$) and \new{that of} the \new{bare} surface ($\alpha_{\rm{surf}}$) as
\begin{equation}
\alpha =1 - (1 - \alpha_{\rm{atm}})(1 - \alpha_{\rm{surf}}) .
\label{eq:albedo}
\end{equation}
\new{In practice, $\alpha_{\rm{atm}}$ is the average of wavelength-dependent scattering efficiency weighted by the spectrum of M-type or G-type stars. The prescription of $\alpha_{\rm{atm}}$ for different types of atmospheres will be detailed below.}

\new{The surface albedo, $\alpha_{\rm{surf}}$, is assumed to be 0.07 for liquid ocean surface and 0.2 for the surface of land planets, regardless of the irradiance spectrum. 
For ocean planets, we also take account of the change of surface albedo due to ocean freezing; when the surface temperature is below 273~K, we assume that the ocean instantaneously freezes and replace the surface albedo by that of ice/snow, which is 0.3 and 0.55 with respect to the spectrum of the M-type star and the G-type star, respectively. The difference in sea ice albedo is due to the redder spectrum of the M-type star where the ice/snow albedo is lower. }

However, the albedo of ocean planets may be \new{better} characterized by water clouds. 
\new{The GCM simulations for} tidally locked ocean planets \new{show that} \new{the region covered by the liquid water on} the dayside is likely to be covered by optically thick water clouds due to convection \citep[e.g.][]{Yang+2013}, while the nightside \new{or frozen surface} tends to be free from thick clouds. In order to take it into account, we modified the albedo for the \new{unfrozen area on the} dayside of ocean planets \new{to 0.4}. 

We summarize our prescriptions of the albedo for each atmospheric condition below
(For the detailed values, see Table \ref{tab:experiment}).

\begin{enumerate}

 \item \new{Ocean planets, Dayside to the M-type star radiation {\it and} above 273~K}:
 
We adopt the cloud-covered albedo \new{0.4}, which is independent of atmospheric composition and pressure.

 \item \new{Ocean planets, Otherwise:}

We adopt the cloud-free albedo given by Eq.~(\ref{eq:albedo}). 

\new{The surface albedo, $\alpha_{\rm surf}$, depends on the surface temperature. Above 273~K, $\alpha _{\rm ocean}=0.07$. Below 273~K, it is 0.3 and 0.55 for the irradiation of the M-type star and the G-type star, respectively. }

\new{The atmospheric albedo with respect to the M-type star spectrum, which is determined by the combination of Rayleigh scattering and atmospheric absorption, is obtained by performing the radiative transfer calculation for each type of atmosphere using SOCRATES \citep{Edwards_Slingo_1996, Edwards_1996} 
described in the appendix \ref{ap:radconv}. A saturated atmosphere with surface temperature of 273~K is assumed.  Precisely speaking, $\alpha_{\rm{atm}}$ depends on the surface temperature due to the change in the column density of water vapor. We also calculated the albedo with the lower surface temperature (200K) and found that deviation in terms of the value of $(1-\alpha)$ is within $\sim10$\%.
The atmospheric albedo with respect to the G-type star are calculated using analytic formula for different types of atmospheres. }

\begin{description}
\item[\rm \underline{type OE}] 
\new{The albedo for M-type star's irradiation is calculated using SOCRATES assuming an 1 bar N2-dominated atmosphere composed of 21\% O2, and 300 ppm CO2.} \new{The albedo for G-type star's irradiation is calculated by the single-scattering approximation with Earth's Rayleigh scattering optical depth by \citet{Young1980}, as described in \citet{Fujii+2010}. }
\item[\rm \underline{type OC}] 
\new{The albedo for M-type star's irradiation is calculated using SOCRATES assuming a pure CO2 atmosphere.} \new{The albedo for G-type star's irradiation is determined based on the analytical expression by \citet{Yokohata+2002} which considered the Martian atmosphere, with a modification due to the difference in gravity (we assume Earth's value for the gravity, $g_{\oplus }$, in this paper):
$\alpha_{\rm atm}=0.021 [\log_{10} (g_{\rm mars}/g_{\oplus })(p/p_{0})] ^{2.5}$ with $p_0 = 6 \times 10^{-3}$ bar. }
\end{description}

\item Land planets:

We adopt the cloud-free albedo \new{given by Eq.~(\ref{eq:albedo}) with the surface albedo is set at 0.2 \citep{Turbet+2016}.
The atmospheric albedo for different types are given as follows:}

\begin{description}
\item[\rm \underline{type LE}] The atmospheric albedo \new{is obtained} by the same calculation as the \new{the cloud-free region} of type OE except \new{that the Rayleigh scattering efficiency is replaced by that of dry air}.

\item [\rm \underline{type LC}] \new{The atmospheric albedo is obtained by} the radiative transfer calculation with SOCRATES (see Appendix \ref{ap:radconv}).
\end{description}

\end{enumerate}

\subsubsection{Thermal emission}  \label{subsubsec:I}

For land planets, the radiation flux $I$ from the top of the planetary atmosphere in Equation (\ref{eq:EBeq}) is given 
in a form of a modified black-body radiation as
\begin{equation}
I_{\rm{land}}= \sigma'(p{\rm{CO}}_{2})  \cdot T^{4}, \label{eq:blarad-CO2}
\end{equation}
where $\sigma'(p{\rm{CO}}_{2})$ is a fitting parameter as a function of CO$_{2}$ partial pressure\footnote{\new{Precisely speaking, $ \sigma'(p{\rm{CO}}_{2})$ also depends weakly on the surface temperature ($T$). In this paper, however, we ignore the dependence for simplicity.}} (Table \ref{tab:experiment}).
With the Earth-like atmosphere, the parameter is approximated by the Stefan-Boltzmann constant in this paper.
With a CO$_{2}$ atmosphere, $\sigma'(p{\rm{CO}}_{2})$ is obtained by our 1D radiative-convective equilibrium calculation. The procedure is detailed in Appendix \ref{ap:radconv}.

For ocean planets, \new{at Earth-like temperatures, $I$ is approximately linear to the temperature} due to the strong greenhouse effect of water vapor \citep[e.g.][and the references therein]{Koll+2018}. \new{Imposing its asymptotic approach to Eq.(\ref{eq:blarad-CO2}) at low temperature, the functional form of $I$ of ocean planets can may written as \citep[][]{Spiegel+2008},
\begin{align}
I_{\rm{ocean}}&= \frac{\sigma'(p{\rm{CO}}_{2}) T^{4}}{1+(3/4)\tau_{\rm IR}}, \label{eq:I_ocean} \\
      \tau_{\rm IR}&=0.79(T/273 {\rm K})^{3}. \label{eq:I_ocean_tau} 
\end{align}
where the coefficient of Eq.~(\ref{eq:I_ocean_tau}) is adopted from \citet{Spiegel+2008}. 
Comparing Eq.~(\ref{eq:I_ocean}) to the linear expression of \citet{Caldeira+1992} which is valid for the range of $10^{-4}$ bar $<$ $p$CO$_2<$ 2 bar and 194K $< T <$ 303K, 
the discrepancy is $\lesssim$10 \% for the most of this range except below 200K.}
Since 3 bar and 10 bar runs are out of this range, we did not calculate these runs for ocean planets.

\new{However, $I_{\rm ocean}$ is also affected by clouds that we assumed for albedo (see section \ref{subsubsec:alb}), as cloud cover tends to reduce the top-of-atmosphere outgoing thermal emission. 
In this paper, we assume the constant cloud-top temperature $T_{\rm cloud-top}=240$~K as a crude approximation referring to the fixed anvil temperature theory \citep{Hartmann+2002} and some GCM results for tidally-locked planets \citep{Yang&Abbot2014}. 
Thus, the thermal emission for the overcast region of the dayside is modified as follows:}
\begin{equation}
I_{\rm ocean, cloudy} = \sigma'(p{\rm{CO}}_{2})  (T_{\rm cloud-top} \simeq 240 {\rm K})^{4}. \label{eq:I_cloud}
\end{equation}

\subsubsection{Diffusion term}  \label{subsubsec:D}
The thermal diffusion \new{due to} the atmospheric and the oceanic flows
can be divided into latitudinal and the longitudinal components:
\begin{align}
\nabla \cdot (\kappa\nabla T(\theta,\phi,t))
=\frac{1}{\cos\theta} \frac{\partial}{\partial \theta}\biggl(D_1(p)\cos\theta\frac{\partial T }{\partial \theta}\biggr)+\frac{1}{\cos^2\theta}\frac{\partial  }{\partial \phi}\biggl(D_2(\theta,p)\frac{\partial T }{\partial \phi}\biggr), \label{eq:diffusionterm}
\end{align}
where $D_1$ and $D_2$ are latitudinal and longitudinal diffusion coefficients, respectively. 
On the Earth, $D_2 \sim 4 D_1$ and their values on the ocean are twice as large as those on the land, which reflects the substantial contribution of oceanic flow to the heat transport.
For ocean and land planets with Earth-like atmospheres, the values for $D_1$ are taken from the Earth's values for ocean and land, respectively \citep{Pollard1983}, while $D_2$ is adjusted for the characteristics of tidally locked planets as follows:
GCM calculations for the tidally locked planets
\citep{Turbet+2016, Turbet+2017, Kopparapu+2017} showed 
characteristic patterns of atmospheric circulation with the coldest regions at high latitudes on the nightside (off the polar regions) associated with \new{the}
zonal flow develop\new{ed} near the equator.
Corresponding to these patterns, \new{ for ocean (land) planets, we set $D_{2}$ to be 0.03 (0.02) times larger than the Earth's value at $\theta > 45$ degrees and to be 1.5 (1.0) times larger than the Earth's value otherwise. }

In order to obtain the values for $D_1$ and $D_2$ for planets with CO$_2$-dominated atmospheres of various surface pressure, we scale the Earth's values assuming the following dependence:
\begin{equation}
D_1, D_2 \propto \frac {p}{g} C_{p},    \label{eq:DDD}
\end{equation}
where $p$ is atmospheric pressure, $g = GM/R^2$, and $C_{p}$ is the heat capacity of the atmosphere. 
We note that the potential dependence on other parameters is ignored here. In reality, atmospheric and oceanic flows that control $D_1$ and $D_2$ would be affected by the spin rate, and irradiation patterns among others. 
\new{For tidally-locked planets, this means $D_1$ and $D_2$ should also depend on the planetary semi-major axis, $a$.} 
The exact dependence of these parameters would be nonlinear, however, and would require the GCM computations. We will discuss this in Section \ref{subsec:beyond}.

\subsubsection{Validation}

\begin{figure*}
\begin{center}
\includegraphics[bb=0.000000 0.000000 465.540000 215.770000, width=0.85\textwidth]{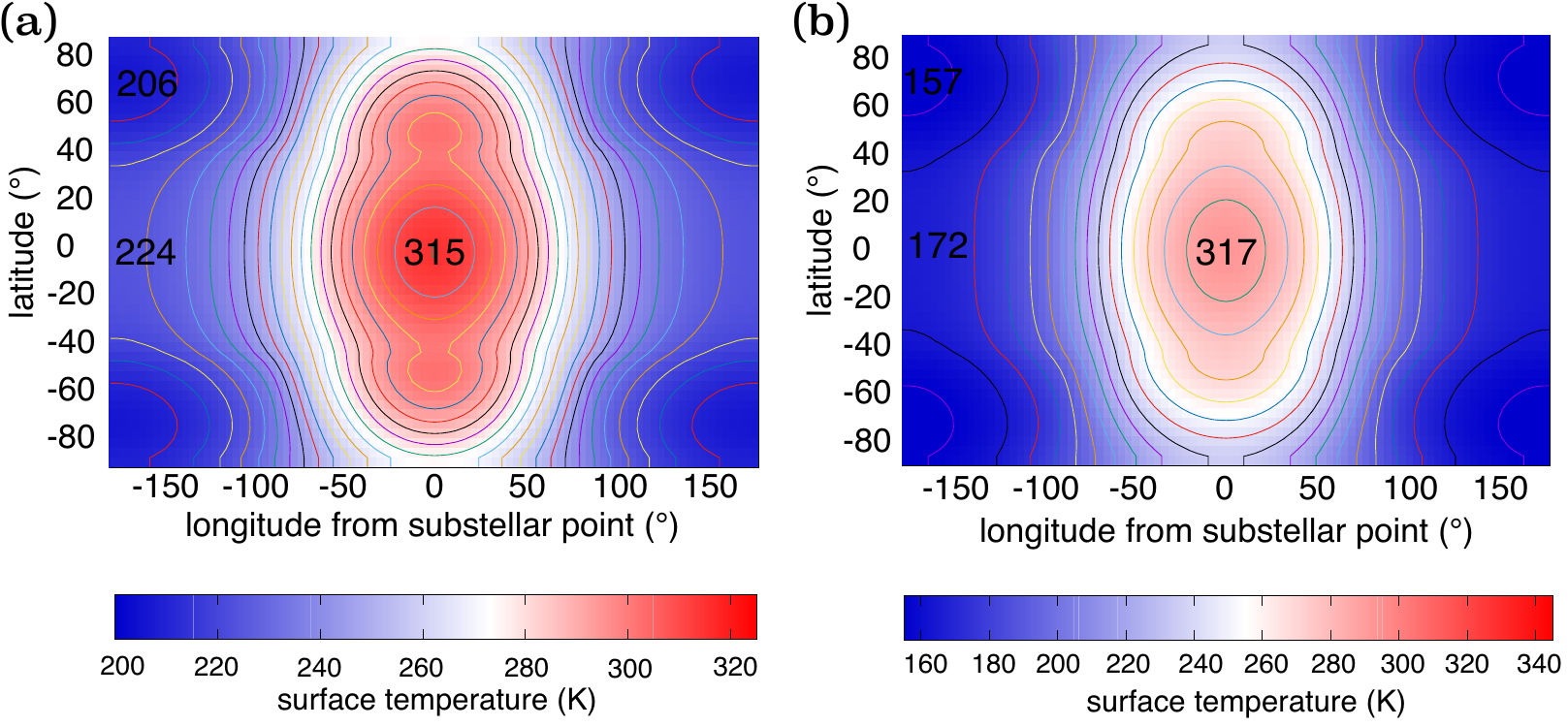}
\caption{The temperature distribution of the Earth-like atmosphere (Table \ref{tab:experiment}) for (a) ocean-covered and (b)  land-covered analogs to Proxima Centauri b.
The horizontal axis shows the longitude from the substellar point of the M-type star. In order to compare (a) and (b) with the Figs. 3 \& 6 in \citet{Turbet+2016} respectively, each color bar for the temperature is set in the same scale as \citet{Turbet+2016}. The map (a) displays contours every 10 K, and the map (b) displays contours every 20 K. }
\end{center}
\label{fig:Proxima}
\end{figure*}

In order to test the validity of our model and parameter setting, we calculated the temperature distribution of Proxima Centauri b, an Earth-size planet
at $a=0.049$ au around a single M-type host star
with $M_* \simeq 0.12 M_{\odot}$ and $L_* \simeq 0.0017L_\odot$.  
Figure \ref{fig:Proxima} shows our result of the 2D distribution of surface temperature for the land and ocean planets with
$g =10.9 \,{\rm m s}^{-2}$ and $F_{\rm{M}}(0,0) = 0.7 S_{\odot}$, where $S_{\odot}$ is the solar irradiation flux at the substellar point, calculated by $M_*, L_*$ and $a$ of the Proxima Centauri system.
The values of the maximum and minimum $T$ and their locations and the overall distribution 
obtained by our model
agree with the previous GCM results for the planet \citep[Figs. 3 and 6,][]{Turbet+2016}.

\begin{table}
\tablenum{1}
\renewcommand{\thetable}{\arabic{table}}
\caption{Model parameters for each experimental condition\label{tab:experiment}}
\begin{splittabular}{cccc|cccBcccc|ccc}
\hline 
\hline 
Surface & \multicolumn2c{Atmosphere} && Heat Capacity & Albedo for the G star's radiation & Albedo for the M star's radiation & Surface & \multicolumn2c{Atmosphere} && Thermal emission & Latitudinal diffusion coefficient &Longitudinal diffusion coefficient \\
\cline{2-3} \cline{9-10}
 & Composition & $P$\ [bar] & &$C$\ [J m$^{-2}$ K$^{-1}$] & ${\alpha_{\rm{G}}}^{*3}$ &${\alpha_{\rm{M}}}^{*4}$ &  & Composition & $P$\ [bar] & & $I$\ [W m$^{-2}$] & $D_{1}$\ [W m$^{-2}$ K$^{-1}$] &$D_{2}$\ [W m$^{-2}$ K$^{-1}]$ \\
\hline 
Ocean & Earth-like & 1 & & $(2.09\times10^{8}, \new{1.67\times10^{7}})^{*1}$ & 
(0.40, \new{0.12, 0.57})  & (0.40,\new{ 0.19}) & Ocean & Earth-like & 1 && $\new{\sigma'=5.67\times 10^{-8}}$ & 0.82 $^{*1}$& \new{4.95} ($\theta<45^{\circ}$) $^{*1}$  \\
Ocean & CO$_{2}$ & 0.3 & & $(2.09\times10^{8}, \new{1.67\times10^{7}})$ & 
(0.40, \new{0.11, 0.57}) & (0.40, \new{0.17})  & Ocean & CO$_{2}$ & 0.3 & & $\new{\sigma'=4.12\times 10^{-8}}$  & 0.21 & \new{1.25} ($\theta<45^{\circ}$) \\
Ocean & CO$_{2}$ & 1 & & $(2.09\times10^{8}, \new{1.67\times10^{7}})$ & 
(0.40, \new{0.16, 0.59}) & (0.40, \new{0.15})  & Ocean & CO$_{2}$ & 1 && $\new{\sigma'=3.19\times 10^{-8}}$  &0.69 & \new{4.15}($\theta<45^{\circ}$) \\
Ocean & CO$_{2}$ & 2 & & $(2.09\times10^{8}, \new{1.67\times10^{7}})$ & 
(0.40, \new{0.20, 0.61})  & (0.40, \new{0.14}) & Ocean & CO$_{2}$ & 2 & & $\new{\sigma'=2.65\times 10^{-8}}$   &1.38 &  \new{8.31}($\theta<45^{\circ}$) \\
\hline
Land & Earth-like & 1 & & ${8.37\times10^{6}}^{*2}$ & \new{0.24} & \new{0.21} & Land & Earth-like & 1 & & 
${\sigma'=5.67\times 10^{-8}} ^{*5}$  & 0.41 $^{*2}$ &\new{1.65} ($\theta<45^{\circ}$) $^{*2}$\\
Land &CO$_{2}$ &0.3 & & ${8.37\times10^{6}}$ & 0.22 & 0.16 & Land &CO$_{2}$ &0.3 & 
& $\sigma'=4.12\times 10^{-8} $ & 0.10 & \new{0.42} ($\theta<45^{\circ}$) \\
Land & CO$_{2}$ & 1 & & ${8.37\times10^{6}}$ & 0.26  & 0.15  & Land & CO$_{2}$ & 1 & &
$\sigma'=3.19\times 10^{-8}$ & 0.34 & \new{1.38} ($\theta<45^{\circ}$)\\
Land &CO$_{2}$ &3 & & ${8.37\times10^{6}}$ & 0.33 & 0.15 & Land &CO$_{2}$ &3 &
& $\sigma'=2.37\times 10^{-8}$ & 1.03 & \new{4.15} ($\theta<45^{\circ}$) \\
Land &CO$_{2}$ &10 & & ${8.37\times10^{6}}$ & 0.44 & 0.17 & Land &CO$_{2}$ &10 &
& $\sigma'=1.78\times 10^{-8} $ & 3.44 & \new{13.8} ($\theta<45^{\circ}$) \\
\hline
\end{splittabular}
\tablecomments{The values (*1) and (*2) refer to those for the ocean and land on Earth being used by \citet{Pollard1983}. \new{The values of $C$ for ocean planets show those for ${T\geq273 \rm{K}}$ and ${T<273 \rm{K}}$.}
\new{The values of albedo for each type of planets are calculated by the prescription in Section \ref{subsubsec:alb}.
$\alpha_{\rm{G}}(*3)$ for ocean planets represents the albedo for cloud-covered, cloudless-unfrozen, and cloudless-frozen areas.
$\alpha_{\rm{M}}(*4)$ for ocean planets represents the albedo for cloud-covered and cloudless-frozen areas.}
The thermal emission of ocean and land planets is given by Eqs.~(\ref{eq:I_ocean}) and (\ref{eq:blarad-CO2}) with a fitting parameter, $\sigma'$\ [W m$^{-2}$ K$^{-4}]$. The value (*5) of the parameter for an Earth-like atmosphere corresponds to the Stefan-Boltzmann constant. 
The diffusion coefficients for CO$_{2}$ atmospheres follow Eq. (\ref{eq:DDD}).}
\end{table}

\subsection{Criteria for temperate climate}\label{subsec:howHZ}

We aim to identify \new{the orbital region where S-type planets with different surface/atmospheric conditions can sustain moderate climates,} as a function of planetary semi-major axis ($a$) and the binary separation ($a_{\rm B}$). 
\new{We focus on the climate with similar temperature range to that Earth experiences, and with the surface liquid water. 
For this, three necessary conditions are considered. 
First is that at least some part of the planet should be above the water freezing temperature. 
Second is that the planet should not undergo atmospheric collapse (otherwise the planet would transition into a much colder state). 
The third one, which is relevant to land planets only, is about the cold trap of water, namely the planet should not confine water to its coldest region in the solid phase. }

\new{In order to discuss these criteria, we use the maximum and minimum temperatures of individual planets at a particular point in time, $T_{\rm{max}}$ and $T_{\rm{min}}$.}
These temperatures oscillate \new{synchronously} to the synodic period between the M-type star and the G-type star relative to the planet, so we \new{can think of} the highest and lowest values of $T_{\rm{max}}$ and $T_{\rm{min}}$ during the synodic period. 
\new{The thresholds corresponding to the three criteria are as follows. }

\paragraph{Water melting}
The first condition, the requirement for the melting water, is expressed by {\it{highest}} $T_{\rm max} > 273$~K (ignoring the minor dependence of freezing temperature on the pressure). 

\paragraph{Atmospheric collapse}
Planets undergo atmospheric collapse if the minimum surface temperature is below the condensation temperature of the major atmospheric composition at least at some point during the synodic period. 
Thus, {\it{lowest}} $T_{\rm min}$ is used to determine if atmospheric collapse should occur.  

\paragraph{Cold trap}
On a land planet, if the coldest region has always had a temperature lower than the freezing point, all water on the planet would eventually be trapped there as permanent ice.
Thus, \new{for land planets to be habitable,} we \new{impose that the {\it{highest}} $T_{\rm{min}}$ to be larger than 273~K. }

We note that it remains unclear if the atmospheric collapse of an ocean planet vitally harms planetary habitability.
Even if the background atmosphere collapses, water vapor that evaporated from the liquid ocean may form a steam atmosphere that allows the planet to retain a habitable condition.
However, water on such planets would not have a long lifetime \citep{Wordsworth+2014}.

\section{Results} \label{sec:results}

In this section, first we discuss the dependence of the surface temperature maps on the binary star separation and the class (ocean or land) of the planet (section \ref{subsec:maps}). Then, we present the planetary orbital radius and the binary star separation that allows for a habitable condition on different types of planets (section \ref{subsec:boundary}). 

\subsection{Global maps of planetary surface temperature} \label{subsec:maps}

In this subsection, we \new{present} the simulated global maps of the planetary surface temperature to show the effects of the irradiation from the G-type star companion.

\begin{figure*}
\centering
\includegraphics[bb= 0.000000 0.000000 527.700000 522.030000, width=0.85\textwidth]{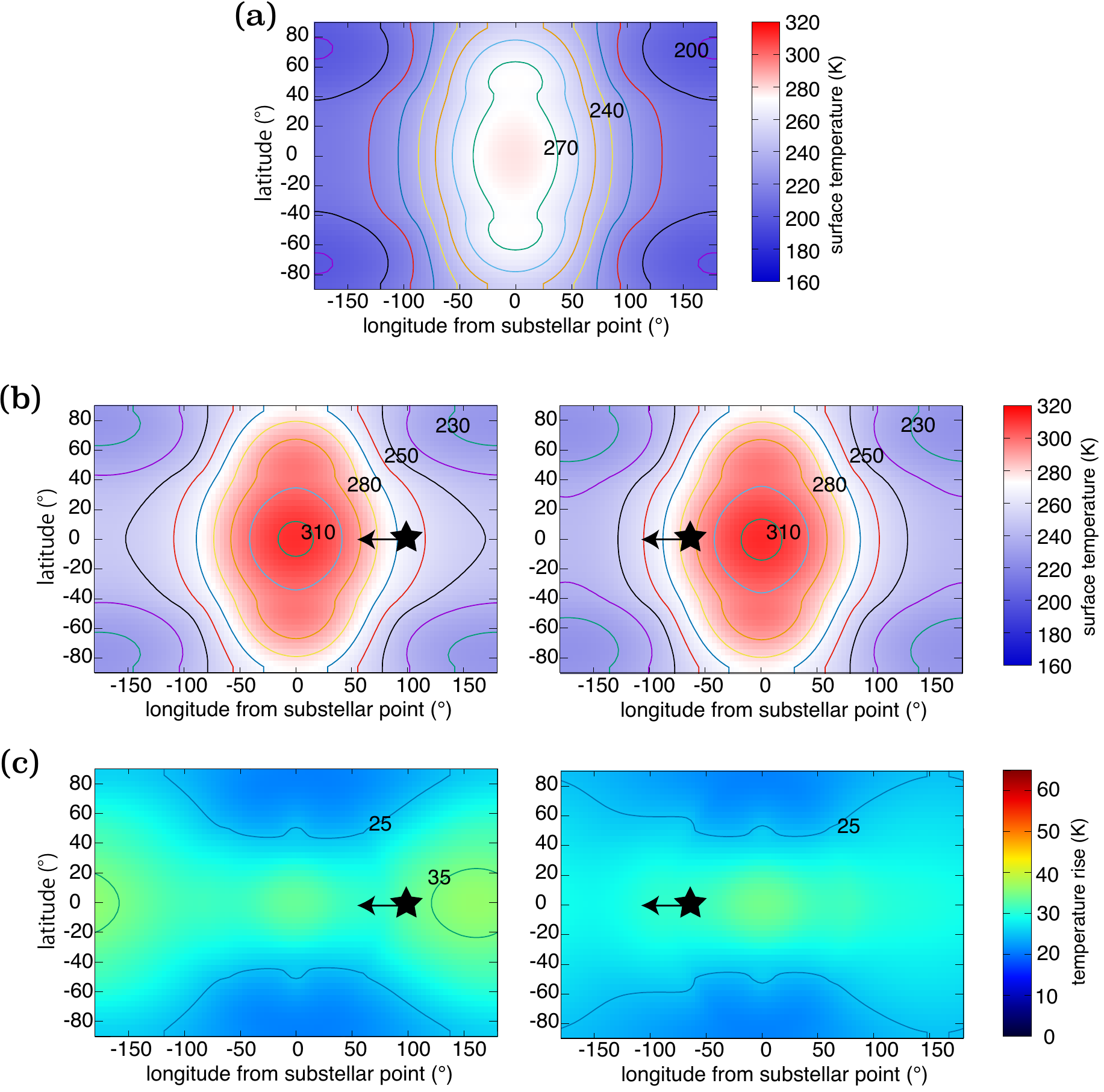}
\caption{
The surface temperature distribution of the ocean planet with a 1-bar Earth-like atmosphere at $a=0.14$ au.
\new{
Panel (a) shows the case around a single M-type star.
Panels (b) is the snapshots in a binary star system with $a_{\rm B} = 1.7$ au.
Panel (c) shows the difference between panels (a) and (b), indicating the temperature change due to the irradiation from the G-type star.}
The star-shaped symbols represent the instantaneous substellar point of the G-type star.
The arrow represents the direction of movement of the substellar points.}
 \label{fig:Tmaps-ocean}
\end{figure*}

\begin{figure*}
\centering
\includegraphics[bb= 0.000000 0.000000 527.250000 521.900000, width=0.85\textwidth]{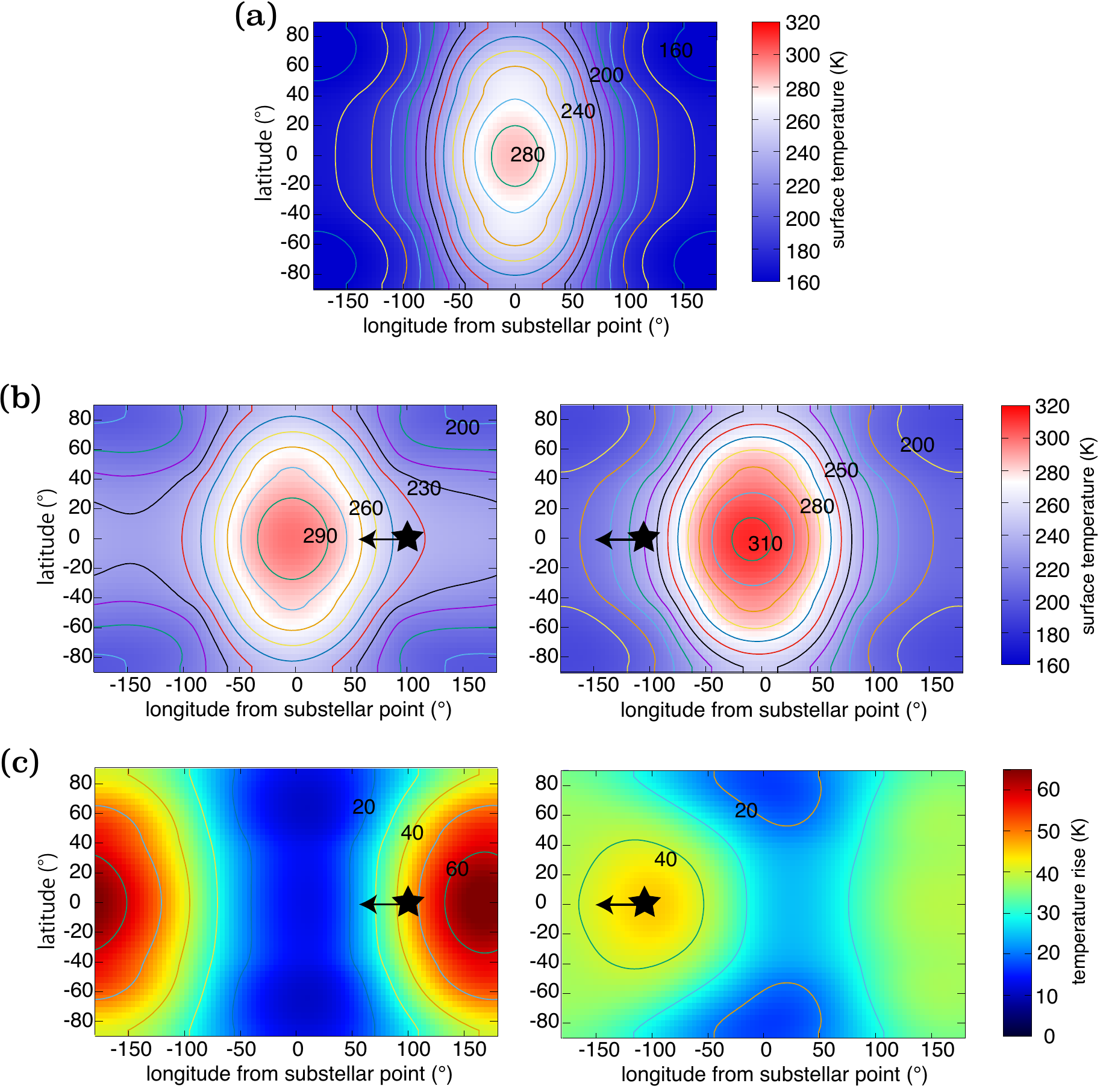}
\caption{
The surface temperature distribution of the land planet with a 1-bar Earth-like atmosphere at $a=0.14$ au.
\new{
Same as panels (a), (b), and (c) in Figure~\ref{fig:Tmaps-ocean} except for the surface environment of the land planet.
}
The star-shaped symbols and arrows are the same as in Figure~\ref{fig:Tmaps-ocean}.
}
 \label{fig:Tmaps-land}
\end{figure*}

Figures \ref{fig:Tmaps-ocean} and \ref{fig:Tmaps-land} show 
the global surface temperature maps for tidally locked ocean and land planets with an Earth-like 1 bar atmosphere, respectively. 
The planets are set at 0.14 au, which is in the classical HZ around a single M-type star \citep{Kopparapu+2013}. 
\new{Panels (a) show the case without } the G-type star irradiation \new{corresponding to} the asymptotic solution for $a_{\rm B}\rightarrow \infty$. 
The central point, $(\phi,\theta)=(0^{\circ},0^{\circ})$, is the M-type star's substellar point. 
\new{Panels (b) show the snapshots of the temperature map in the case of S-type planets with a G-type companion star at $a_{\rm B} = 1.7 \,{\rm au}$. The G-type star's substellar point is indicated by the star symbol. }
\new{We note that for ocean planets we carried out two calculations with the different initial conditions, one with globally freezing temperature and the other with globally melted temperature and confirmed that the results are same. }

\new{On average, with a given orbital configuration, land planets are colder than the ocean planets because of the lack of the greenhouse effect caused by water vapor. 
In addition, land planets have larger temperature gradient than ocean planets because of the smaller diffusion coefficients. 
Therefore, the minimum temperature of land planets is in general colder than ocean planets, which suggests that it is easier for land planets to undergo atmospheric collapse or the cold trap of water.}

\new{The bottom panels show the difference between panels (a) and (b), indicating the temperature change due to the irradiation from the G-type star. }
\new{In the case of ocean planets (Figure~\ref{fig:Tmaps-ocean}), } we found that the temperature increase by the G-type star's irradiation is
zonal and almost independent of time; equivalently, it is also independent of the location of
the G-type star's substellar point.
The maximum temperature is always located at
the substellar point of the M-type star.
\new{This feature is} explained by comparison of timescales: the synodic period,
the thermal relaxation timescale, 
and the latitudinal and longitudinal thermal diffusion timescales.
From Eq.~(\ref{eq:omega}), the synodic period is
\begin{eqnarray}
t_{\rm syn} & = & \frac{2\pi}{\sqrt{GM_{\rm M}/a^3} - \sqrt{G(M_{\rm G}+M_{\rm M})/a_{\rm B}^3}} \\
 & = & 0.105 \left[ \left(\frac{M_{\rm M}}{0.25 M_{\odot}}\right)^{1/2} 
\left(\frac{a}{0.14\,{\rm au}}\right)^{-3/2}
- \new{0.05} \left(\frac{M_{\rm G}+M_{\rm M}}{1.25 M_{\odot}}\right)^{1/2} \left(\frac{a_{\rm B}}{\new{1.7}\,{\rm au}}\right)^{-3/2}\right]^{-1}{\rm yr}.
\label{eq:t_syn}
\end{eqnarray}
The thermal relaxation (response) time for ocean planets is estimated as follows, based on
Eqs.~(\ref{eq:EBeq}), (\ref{eq:I_ocean}), and Table 3: 
\begin{eqnarray}
\new{
t_{\rm relax}} & \sim & \frac{C T}{I}  \simeq 
   \left\{
   \begin{array}{cl}
   \frac{C_{\rm ocean}T}{\sigma {T^4}_{\rm cloud\mathchar`-top}}=10.6 \left(\frac{T}{300\,{\rm K}}\right) \,{\rm yr} & \; [{\rm cloud\mathchar`-covered},\ T > 273K] \\
   \frac{C}{\sigma T^{3}} \times \frac{3}{4} \tau_{\rm IR} \sim 3.4 \left(\frac{C}{C_{\rm ocean}}\right) \,{\rm yr} & \;[{\rm cloud\mathchar`-free},\ T \sim 273 K] \\
   \frac{C_{\rm ice}}{\sigma T^3} \simeq 0.5 \left(\frac{T}{200\,{\rm K}}\right)^{-3} \,{\rm yr} & \; [{\rm cloud\mathchar`-free},\ T \ll 273K].
 
   \end{array} \label{eq:t_relax_ocean}
   \right.
\end{eqnarray}
From Eqs.~(\ref{eq:EBeq}), (\ref{eq:diffusionterm}), and Table 3, the latitudinal and longitudinal thermal diffusion timescales are 
\begin{eqnarray}
t_{\rm diff,\theta} & \sim & \frac{C}{D_1} \simeq  \sim 8.1 \left(\frac{C}{C_{\rm ocean}}\right) \,{\rm yr}, \label{eq:t_diff_theta}\\
t_{\rm diff,\phi} & \sim & \frac{C \cos^2 \theta}{D_2} \simeq 
   \left\{
   \begin{array}{cl}
   1.8 \left(\frac{C}{C_{\rm ocean}}\right) \cos^2 \theta \;{\rm yr} & \; [\theta < 45^\circ] \\
   101 \left(\frac{C}{C_{\rm ocean}}\right) \cos^2 \theta \;{\rm yr} & \; [\theta > 45^\circ],
   \end{array} \label{eq:t_diff_phi}
   \right.
\end{eqnarray}
\new{Thus, for ocean planets, $ t_{\rm relax} \sim t_{\rm diff,\theta} > t_{\rm syn}$.}
\new{As indicated by the panels (a) and (b)}, the temperature distribution reflects the asymmetry of the irradiation from the M-type star both latitudinally and longitudinally, \new{and this is consistent with}
 $t_{\rm relax} \sim t_{\rm diff}$. 
\new{However, because $t_{\rm syn} < t_{\rm relax}$, } the temperature distribution contributed by the G-type star is \new{longitudinally avaeraged}.
As a result, the minimum temperature is \new{found} not at the antistellar point, but 
at \new{the high-latitude region} on the nightside area while the maximum temperature is always at the substellar point of the M-type star.

In contrast, the temperature distribution on the land planet 
follows the time variation of the substellar point of the G-type star, as shown in Figure~\ref{fig:Tmaps-land}.
\new{The main reason is that} the heat capacity is much lower for the land planets.
 \new{(} $C = 8.37\times 10^6 \,{\rm Jm^{-2}K^{-1}}$ for the land planets, 
while $C = 2.09\times 10^8 \,{\rm Jm^{-2}K^{-1}}$ for the ocean planets\new{)}.
\new{Adopting the values for land planets with Earth-like atmospheres, the thermal relaxation timescale for land planets is}
\begin{eqnarray}
t_{\rm relax} \sim \frac{C T}{I} \sim \frac{C}{\sigma T^3} \simeq 0.15 \left(\frac{T}{300\,{\rm K}}\right)^{-3} \,{\rm yr},
\label{eq:t_relax_ocean}
\end{eqnarray}
\new{while the diffusion timescale is} 
\begin{eqnarray}
t_{\rm diff,\theta} & \sim & \frac{C}{D_1} \simeq  \frac{ 8.37\times 10^6}{0.41} \,{\rm s} \sim 0.65 \,{\rm yr}, \\
t_{\rm diff,\phi} & \sim & \frac{C \cos^2 \theta}{D_2} \simeq 
   \left\{
   \begin{array}{cl}
   0.14\cos^2 \theta \;{\rm yr} & \; [\theta < 45^\circ] \\
   8.1 \cos^2 \theta \;{\rm yr} & \; [\theta > 45^\circ],
   \end{array}
   \right.
\end{eqnarray}
\new{Like} the case of land planets, $t_{\rm relax} \sim t_{\rm diff}$,
and \new{the effect of the distribution of the irradiation} is preserved. 
\new{On the top of it, because $t_{\rm relax} \sim t_{\rm syn}$, the contribution of the G-type companion star is not longitudinally averaged, and the temperature map traces its location with a slight delay.}

\begin{figure}[ht!]
\begin{center}
   \includegraphics[bb=0.000000 0.000000 269.700000 163.170000,width=0.6\hsize]{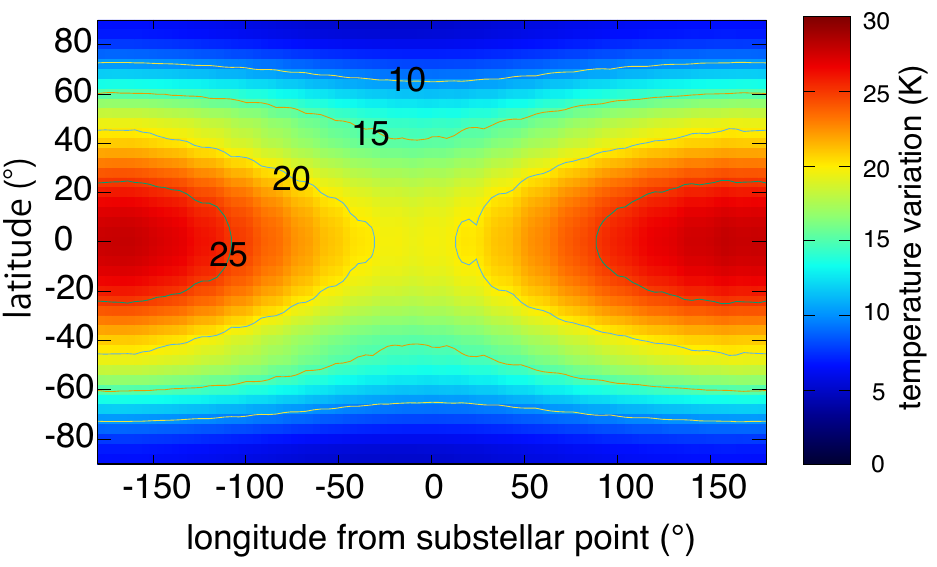}
  \end{center}
\caption{The global map of temperature variation during one synodic period for a land planet with 0.14 au orbital radius and 1.7 au binary separation. The antistellar point of the M-type star ($\theta=0^{\circ}$ and $\phi=180^{\circ}$) has the largest amplitude of the time variation of temperature.}
\label{fig:deltaTmap}
\end{figure}

\new{The temperature maps of land planets, Figure~\ref{fig:Tmaps-land}, also indicate that the increase in temperature due to the G-type star irradiation is more significant on the nightside. 
This can also be seen in Figure \ref{fig:deltaTmap} which presents the amplitude of the time variation of local temperature in the case of Figure~\ref{fig:Tmaps-land}. 
This is because the point corresponds to the substellar point of the G-type star when the G-type star is the closest to the planet (i.e., at the conjunction).} \new{In fact,} this trend is independent of the binary separation, orbital radius, planetary surface (land-covered/ocean-covered), and the atmospheric compositions and pressure.
However, it is much more pronounced for the land planets\new{, because of the regionally confined effect of the G-type companion star}.

\subsection{Effects of the Companion Star on $T_{\rm max}$ and $T_{\rm min}$} \label{subsec:profile}

\begin{figure*}
 \begin{minipage}{0.5\hsize}
  \begin{center}
   \includegraphics[bb= 0.000000 0.000000 377.440000 292.960000,width=1\textwidth]{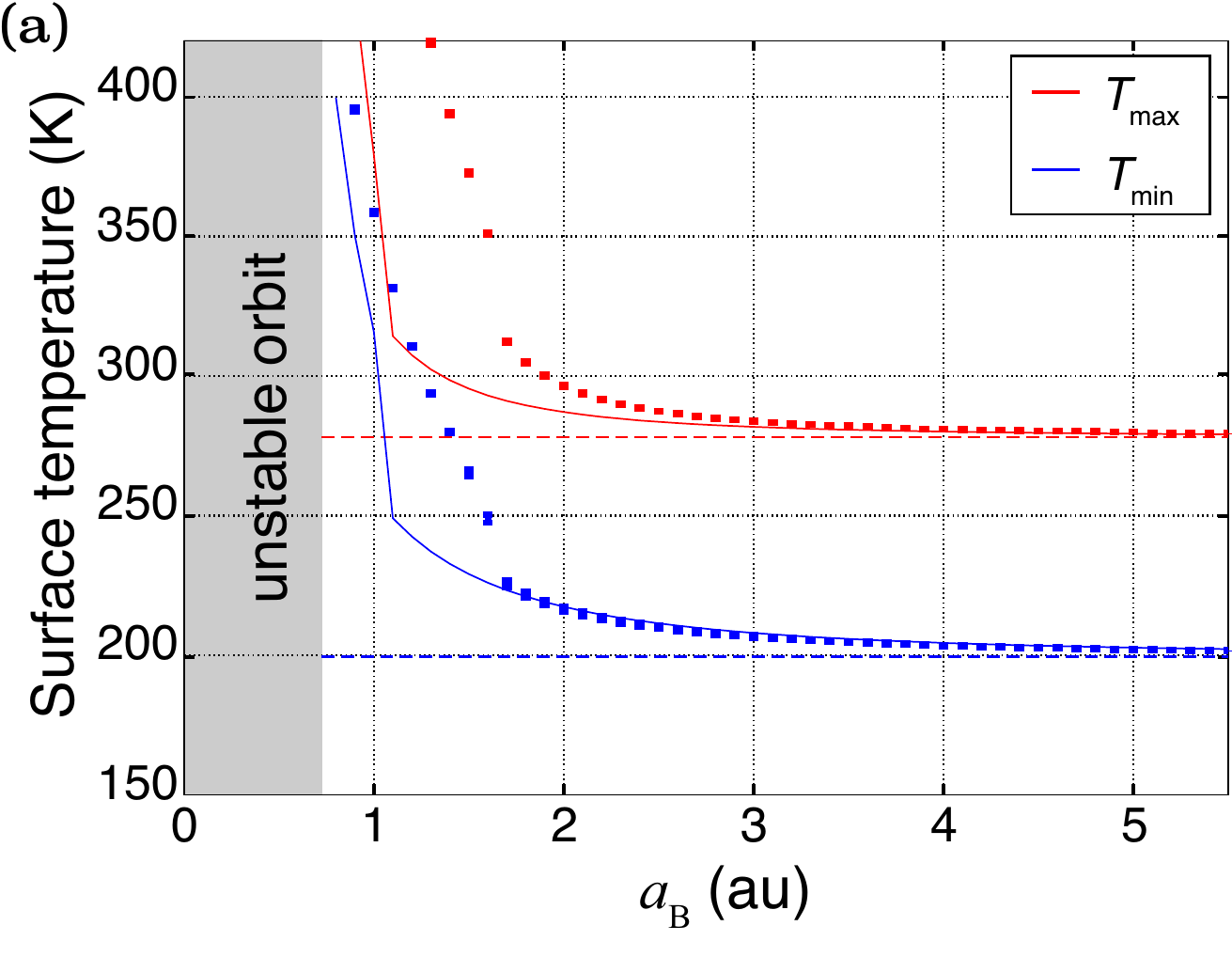}
  \end{center}
 \end{minipage}
 \begin{minipage}{0.5\hsize}
  \begin{center}
   \includegraphics[bb=0.000000 0.000000 377.440000 292.960000,width=1\textwidth]{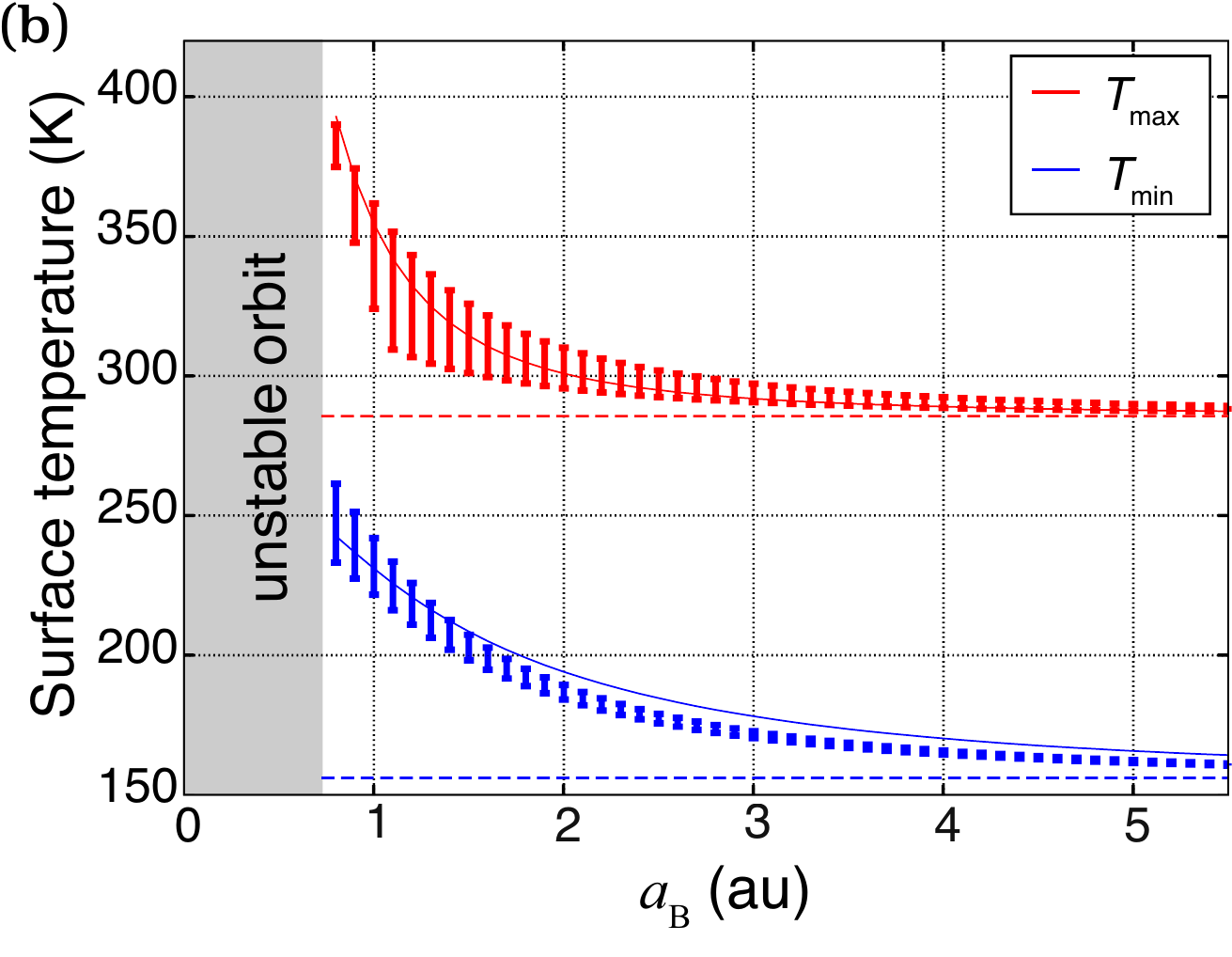}
  \end{center}
 \end{minipage}
\caption{The planetary surface temperature profile of (a) the ocean planets and (b) the land planets as 
a function of the binary separations $a_{\rm B}$ for a fixed planetary semimajor axis $a = 0.14$ au. 
The red and blue plots are the maximum and minimum temperatures, respectively.
The bars show the temporal variation in one synodic period. 
The dotted lines show the temperature of a single M-type star case. 
The solid lines in (a) and (b) show the approximate value given by Eq.~(\ref{eq:landapp}).}
\label{fig:3point}
\end{figure*}

\new{In order to see closely the effect of the G-type star on the potential habitability of the planets,}
we plot the maximum and minimum temperatures ($T_{\rm max}$ and $T_{\rm min}$), \new{our measures of planetary climate,} as a function of $a_{\rm B}$
for $a=0.14$ au in Figure~\ref{fig:3point}.
The left and right panels show the results of ocean planets and land planets, respectively.
We plot the ranges of the variations of 
$T_{\rm max}$ and $T_{\rm min}$ by the vertical bars.
While the temperature distribution on the ocean planets are almost time-independent,
that on the land planets varies in the synodic time\new{, as we discussed in the previous section.}.
In the \new{shaded} region, 
the planetary orbit is destabilized by the companion star's perturbations (Eq.~\ref{eq:stableorbit}).
The asymptotic values of $T_{\rm max}$ and $T_{\rm min}$ for large $a_{\rm B}$
correspond to the result of a single M-type star case.

As $a_{\rm B}$ decreases, both $T_{\rm max}$ and $T_{\rm min}$ are raised by the irradiation from the G-type companion star, and it becomes substantial around 1 au or smaller. 
\new{At large $a_B$,} we found that the temperature distribution in the binary system $T_{\rm MG}(\theta, \phi )$ is approximately given by
\begin{equation}
T_{\rm MG}^{4} (\theta, \phi ) \simeq T_{\rm M}^{4}(\theta, \phi ) + T_{\rm G}^{4}(\theta, \phi ). \label{eq:landapp}
\end{equation}
where $T_{\rm M}(\theta, \phi )$ and $T_{\rm G}(\theta, \phi )$ are the temperature distribution
with only the M-type star and that with only the G-type star. 
\new{Thus, when the G-type companion star is around a few au, the temperature increase is $\sim$ 10 K.}
Although the temperature increase by as small as $\sim$ 10 K appears trivial, such a small change can actually have impacts on the \new{habitable condition} in some cases, because the baseline temperature of the nightside determined by the irradiation from the M-type star is not far from the condensation temperatures of atmospheric constituents. 
A small addition to this baseline temperature can therefore save the planet from atmospheric collapse. 
\new{This will be discussed further in section \ref{subsec:boundary} below.}

\new{We note that, in the case of ocean planets, the effect of the G-type stars' irradiation on $T_{\rm min}$ at large distance would be much larger if the increase in planetary albedo due to ocean freezing were not taken into account. 
Once the ocean starts to freeze on the nightside, the increased planetary albedo with respect of the G-type star limits the effect of the G-type companion at large separation. }

\new{In the previous section, we have discussed that the temperature increase of land planets  measured by $\Delta  T_{\rm MG} \equiv T_{\rm MG} -T_{\rm M}$ is larger on the nightside (Panel (c) of Figure~\ref{fig:Tmaps-land}). 
Consistently, the right panel of Figure~\ref{fig:3point} indicates that the increase in $T_{\rm min}$ due to the companion star is larger than that of $T_{\rm max}$ at any orbital configuration.  
The fractional increase is even more pronounced for $T_{\rm min}$, because $T_{\rm min}$ would be very small without a companion star. 
In other words, $T_{\rm min}$ is more sensitive to $a_B$ than $T_{\rm max}$ is. 
This will be one of the key features that affect the planetary climate as a function of orbital parameters, which will be discussed in the next section.}

\subsection{\new{Orbital region for temperate climate}} \label{subsec:boundary}

Now, we perform similar EBM simulations changing both $a_B$ and $a$ with intervals of $(\Delta a_{\rm B}, \Delta a) = (0.1\;{\rm au}, 0.02 \;{\rm au})$ to produce contour maps
of $T_{\rm max}$ and $T_{\rm min}$ on the $a_{\rm B}-a$ plane. 
\new{Taking account of} our thresholds of $T_{\rm max}$ and $T_{\rm min}$ for  \new{the water cold trap and atmospheric collapse (Section \ref{subsec:howHZ})},
we discuss \new{the region on the $a_{\rm B}-a$ plane where the planetary climate is temperate and allows for the surface liquid water}. 
In the following, we discuss ocean planets and land planets separately. 

\subsubsection{Ocean planet} \label{subsubsec:ocean}
\begin{figure*}
\begin{center}
\includegraphics[bb=0.000000 0.000000 281.470000 163.810000, width=0.6\textwidth]{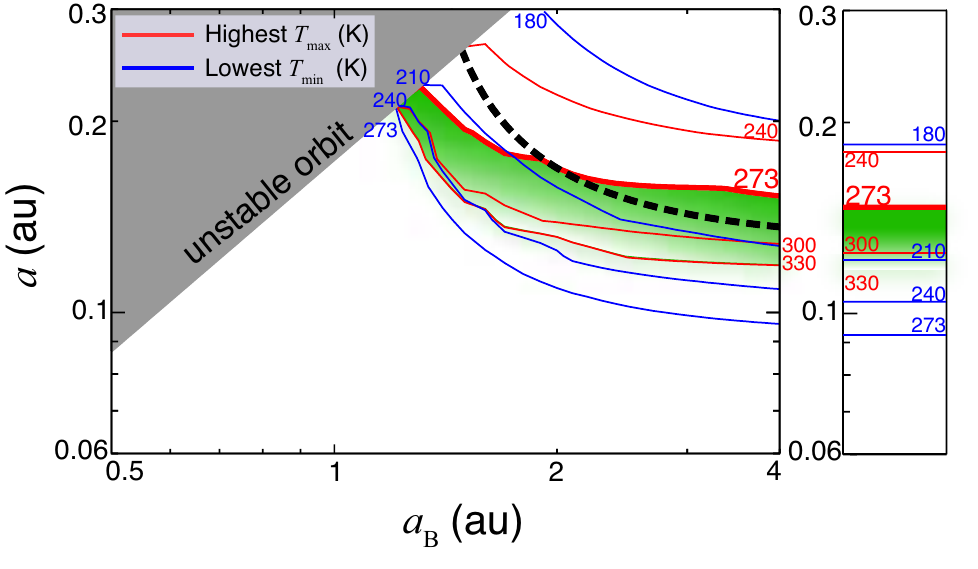}
\caption{\new{The areas (green areas) where an ocean planet with an Earth-like atmosphere can maintain Earth-like mild climate} in a binary system (left) compared with single M star case (right). Both vertical lines show the semimajor axis $a$ and, in the left figure, the horizontal line shows binary separation $a_{\rm B}$. The red and blue contours are $T_{\rm max}$ and $T_{\rm min}$, respectively. 
\new{The reference inner boundary corresponds to the maximum temperature of present-day Earth., $T_{\rm max} \sim 330$ K.} The outer boundary corresponds to the ocean freezing line, $T_{\rm max} = 273$ K. 
The shaded area is the orbitally unstable region (Eq.~\ref{eq:stableorbit}).
\new{The black dotted line shows the orbital configuration for which combined averaged incoming stellar flux is 0.6$S_{\odot}$.}
}
\end{center}
\label{fig:BoundaryOcean}
\end{figure*}

 \new{The left panel of Figure~\ref{fig:BoundaryOcean} shows the contours of $T_{\rm max}$ and $T_{\rm min}$ for S-type ocean planets with Earth-like atmospheres.}
The narrow right box represents \new{those} around a single M-type star, which is the asymptotic solution for $a_{\rm B}\rightarrow \infty$ \new{(the horizontal axis in the narrow right box is a dummy for presentation purposes)}. 
\new{The minor non-smoothness of the contours from $T_{\rm min}\geq 273$ K to  $T_{\rm max}\leq 273$ K arises from the numerical instability due to the discontinuities in the albedo and thermal emission parameterizations (see section \ref{subsec:EBM}). }

\new{The orbital region where the planets can have Earth-like moderate climate is shown in graded green.} 
\new{Here,} the outer boundary is \new{determined by} the requirement that the ocean has to be ice-free at least at some point, $T_{\rm max} > 273 {\rm K}$. 
Although atmospheric collapse should also be taken into account, the major atmospheric component of an Earth-like atmosphere, N$_2$, only condenses at $T<$ 79K, which is much lower than the minimum temperature here and this does not interfere with the \new{areas with moderate climate} found above. 
\new{The orbital region that satisfies the above criterion ($T_{\rm max} > 273 {\rm K}$) is colored as far as $T_{\rm max} \lesssim 330 {\rm K}$ in reference to the maximum temperature of present-day Earth, and this region is approximately considered as having temperate climate. 
In reality, it is likely that the climate of ocean planets is destabilized into the runaway greenhouse regime at certain irradiation level. 
However, the exact location of this threshold for these S-type planets are not known and cannot be determined within the framework of EBM. 
We express the uncertainty by the gradation in color and discuss this uncertainty in Section \ref{subsec:tidally-HZ} below. }

In the case of the single M-type star, the inner and outer
radii of the \new{area with Earth-like temperate temperatures} are $a_{\rm in}= 0.12$ au and $a_{\rm out}= 0.14$ au, respectively.
As $a_{\rm B}$ becomes smaller, both $a_{\rm in}$ and $a_{\rm out}$ 
are increased by the irradiation from the G-type star.
For $a_{\rm B} < 1.3$ au, the \new{temperate area} overlaps the orbitally unstable region
and the width \new{of the area} effectively becomes smaller.

\begin{figure*}
\begin{center}
\vspace{0.5cm}
\includegraphics[bb=0.000000 0.000000 285.630000 170.380000, width=0.63\textwidth]{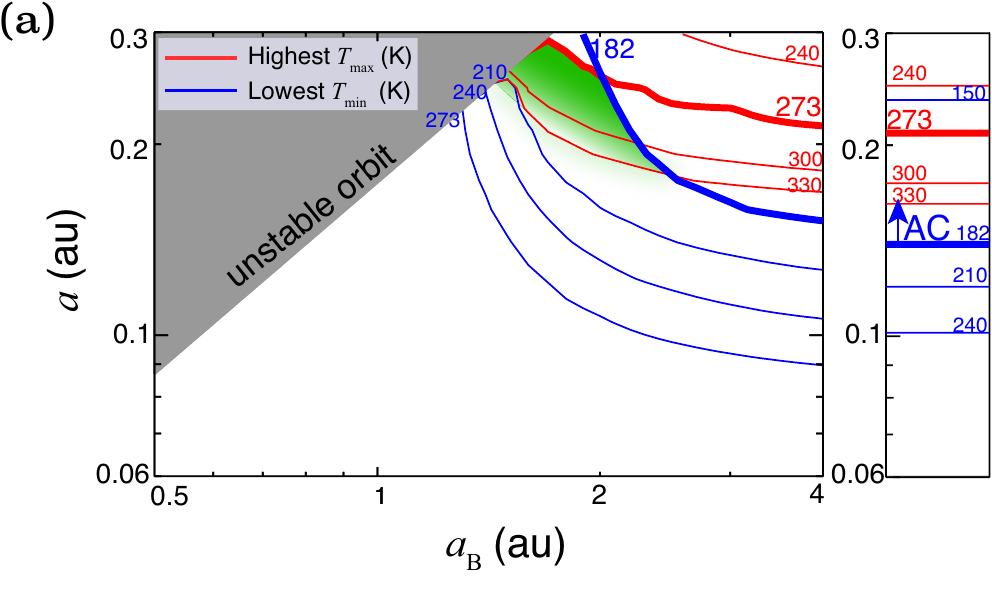}
\includegraphics[bb=0.000000 0.000000 293.480000 169.900000, width=0.63\textwidth]{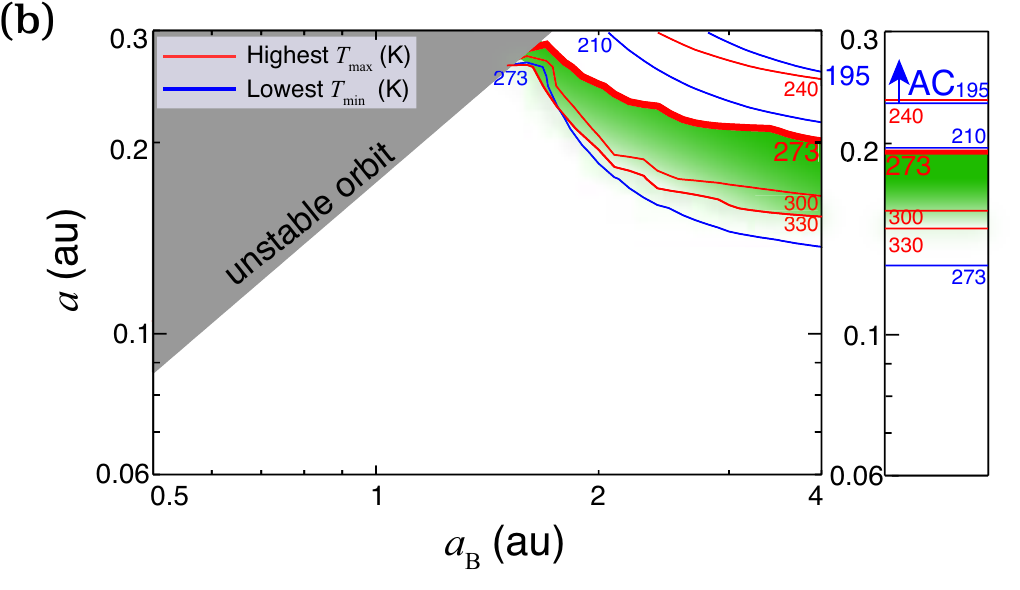}
\vspace{0.5cm}
\includegraphics[bb=0.000000 0.000000 294.070000 170.750000, width=0.63\textwidth]{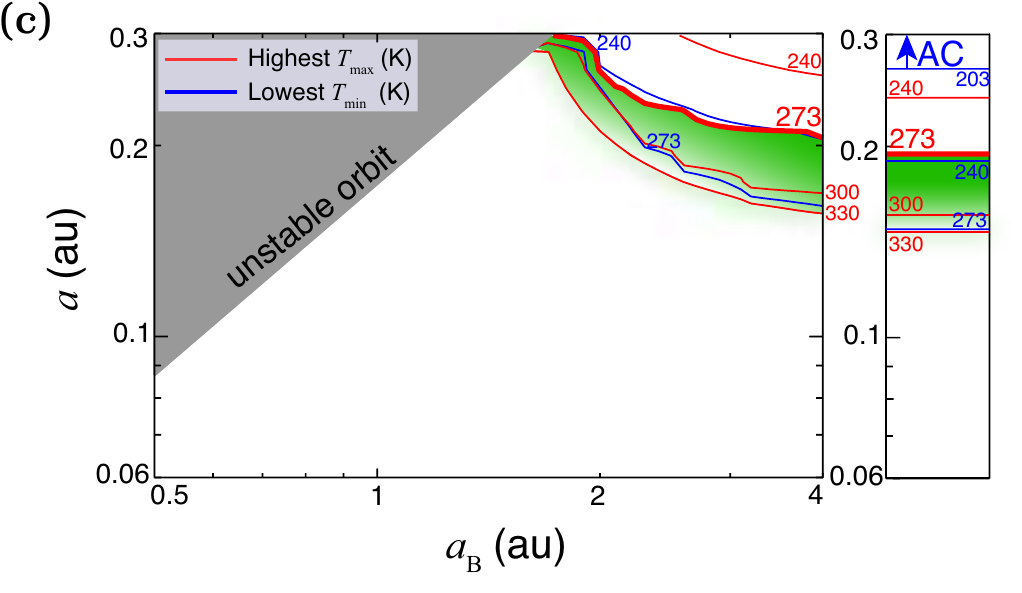}
\caption{Same as Figure~\ref{fig:BoundaryOcean} except for
CO$_2$ atmosphere with (a) 0.3, (b) 1, (c) 2 bar. 
The AC line stands for Atmospheric Collapse and it is determined
by $T_{\rm min}$ of the CO$_{2}$ freezing point, (a) 182 K, (b) 195 K, and (c) 203 K, respectively.  }
\end{center}
\label{fig:BoundaryOceanCO2}
\end{figure*}

Figure ~\ref{fig:BoundaryOceanCO2} shows similar contour maps but with CO$_{2}$ atmospheres of varying surface pressures. 
In this case, while the trend in $T_{\rm max}$ remains, atmospheric collapse becomes an important factor. 
The CO$_{2}$ condensation temperature is 182, 195, and 203 K for $p{\rm CO}_{2} = 0.3, 1$, and 2 bar, respectively. If $T_{\rm min}$ is lower than these values, the atmospheric CO$_2$ starts to condense out to the surface.
As shown in the right panels of Figure~\ref{fig:BoundaryOceanCO2}, without a companion star, ocean planets with $<1$ bar CO$_{2}$ atmosphere that would otherwise have a habitable range of $T_{\rm max}$ ($273\;{\rm K} < T_{\rm max} < 330\;{\rm K}$) cannot avoid atmospheric collapse on the nightside, due to the large temperature gradient between the dayside and the nightside. 
In the binary system, the irradiation from the G-type star raises the nightside temperature and can rescue the planet from atmospheric collapse. 
The companion star at a distance of \new{2.5 au} can produce \new{areas with mild climate} which would otherwise be nonexistent (Figure~\ref{fig:BoundaryOceanCO2} (a)). 
Although the irradiation from the G-type star is weak at such a distance, the small addition of the heat to the nightside greatly contributes toward raising the cold nightside temperature while the dayside temperature only changes a little, allowing for \new{the temperate climate}.

\new{The key for the emergence of the \new{orbit with the habitable climate} is the higher sensitivity of the constant-$T_{\rm min}$ lines to the binary separation than the constant-$T_{\rm max}$ lines. 
The reasons are two-fold. 
As a baseline, the dependence of $T_{\rm min}$ on $a$ in the absence of the G-type companion star is weaker than $T_{\rm max}$. 
On the top of it, the companion star has larger effect on $T_{\rm min}$ than on $T_{\rm min}$, as discussed in Section \ref{subsec:profile}. 
Therefore, the constant-$T_{\rm min}$ lines are strongly skewed by $a_{\rm B}$, while the constant-$T_{\rm max}$ lines are closer to the constant-a lines. 
These are general outcome of our climate modeling.} 

\subsubsection{Land planet} \label{subsec:land}

\new{When producing similar contour maps for land planets, we have to be aware that } $T_{\rm max}$ and $T_{\rm min}$ of the land planets
change \new{during} the synodic period
\new{Here, we plot their {\it highest} values because they are more relevant to the evaluation for the temperate climate, for the reasons described below. }

The outer boundary \new{of the area with habitable climate} is determined by either the cold trap and atmospheric collapse, but based on the simulations in Section \ref{subsec:maps}, we can see that  the cold trap is very likely to be more severe in the case of land planets, for the following reason: 
we assume that the cold trap occurs if $T_{\rm min}$ is always $< 273$ K
during the synodic period, which means the limit is where the {\it highest} value of $T_{\rm min}$ is equal to $273$ K. 
On the other hand, we assume the atmospheric collapse occurs 
if there is a moment at which $T_{\rm min}$ is lower than the CO$_2$ condensation temperature
during the synodic period, that is, the limit is where the {\it lowest} value of $T_{\rm min}$ is equal to the condensation temperature of the atmospheric consituents--- N$_2$ is 79 K for the 1-bar Earth-like atmosphere, and 182--233 K for 0.3--10 bar ${\rm CO}_2$ atmospheres. 
Figure \ref{fig:3point} (b) shows that the variation amplitude of $T_{\rm min}$ is as small as $\sim $ 10--20 K. 
Because the condensation temperature is significantly lower than 273 K, it is cold trap that actually determines the outer boundary of the \new{habitable climate area}. 
After all, the {\it highest} value of $T_{\rm min}$ is important for the outer boundary.

\new{On the other hand, we assume the same inner limit as that of ocean planets, which is $T_{\rm max} \leq 330$K, to identify the area with Earth-like mild climate.
In this section, $T_{\rm max}$ represents its highest value during the synodic period.}

\begin{figure*}
\begin{center}
\includegraphics[bb=0.000000 0.000000 283.820000 163.410000, width=0.6\textwidth]{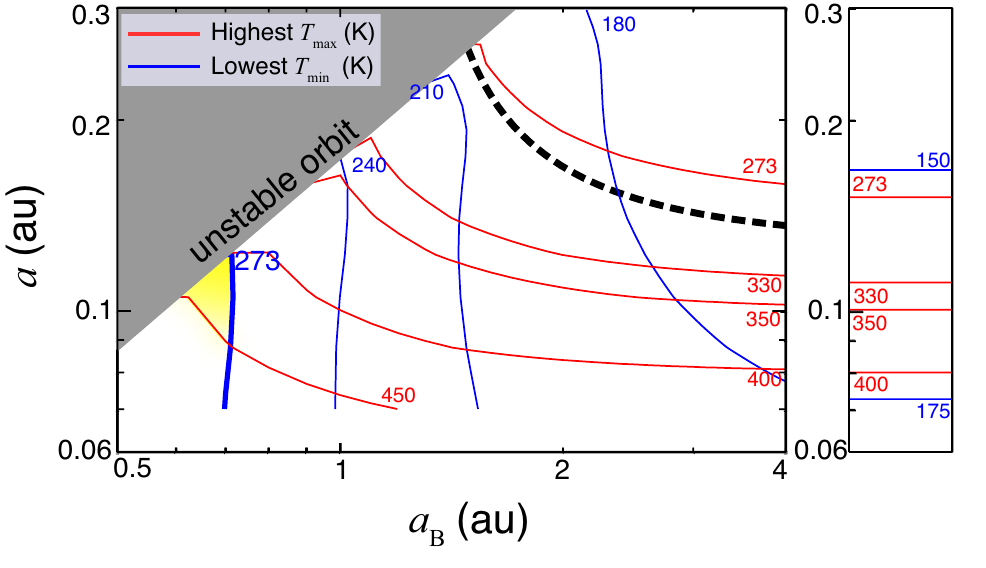}
\caption{\new{The area with habitable climate including the area with Earth-like mild climate (green areas) and potentially habitable areas (yellow areas)}
as a function of the binary separation $a_{\rm B}$ and
the planetary semimajor axis $a$ for land planets with 
Earth-like atmospheres in binary systems (the left panels).
The right panel shows the single M-type star cases for comparison. 
The red and blue contours are $T_{\rm max}$ and $T_{\rm min}$, respectively. \new{While the reference inner boundary for the green area is set by $T_{\rm min} \sim 330$ K, that for the yellow area is  $T_{\rm max} \sim 450$ K. The outer boundary of the both areas is defined by $T_{\rm min} = 273$ K. 
}}
\end{center}
\label{fig:BoundaryLand}
\end{figure*}

The contours of (highest) $T_{\rm max}$ and $T_{\rm min}$ for land planets with Earth-like atmospheres in a binary system are presented in the left panel of Figure \ref{fig:BoundaryLand}, while that of the planets around a single M-type star are shown on the right.
The planet around a single M-type star does not have \new{an area of Earth-like temperatures} because of the extreme day-night temperature difference.

\new{In the case of a binary system, } the irradiation from the G-type star drastically heats up the nightside,
and the water freezing line in the planetary $a$ drastically increases, while the dayside temperature is still dominated by the M-type star. 
\new{
As a result, there is a orbital region where the nightside is warm enough to avoid the cold trap or the atmospheric collapse, while the dayside temeprature is about 400~K or less. 
Such planets have extreme climate beyond the range the Earth experiences, and it is not clear whether such planets can be called habitable. 
However, it should be noted that the climatological transition of the land planets into the runaway greenhouse state is likely to occur at much higher $T_{\rm max}$ than ocean planets \citep[e.g.,][]{Abe+2011, Kodama+2018} (see section \ref{subsec:tidally-HZ}), because land planets can emit thermal energy to space from the hottest region. 
Considering this possibility, we show the extended potentially habitable area in yellow area in Figure~\ref{fig:BoundaryLand}.
The area appears only in the binary system, when the G-type companion star is at $\sim 0.7$ au. 
}

\begin{figure*}
\begin{center}
\includegraphics[bb=0.000000 0.000000 284.830000 170.400000, width=0.51\textwidth]{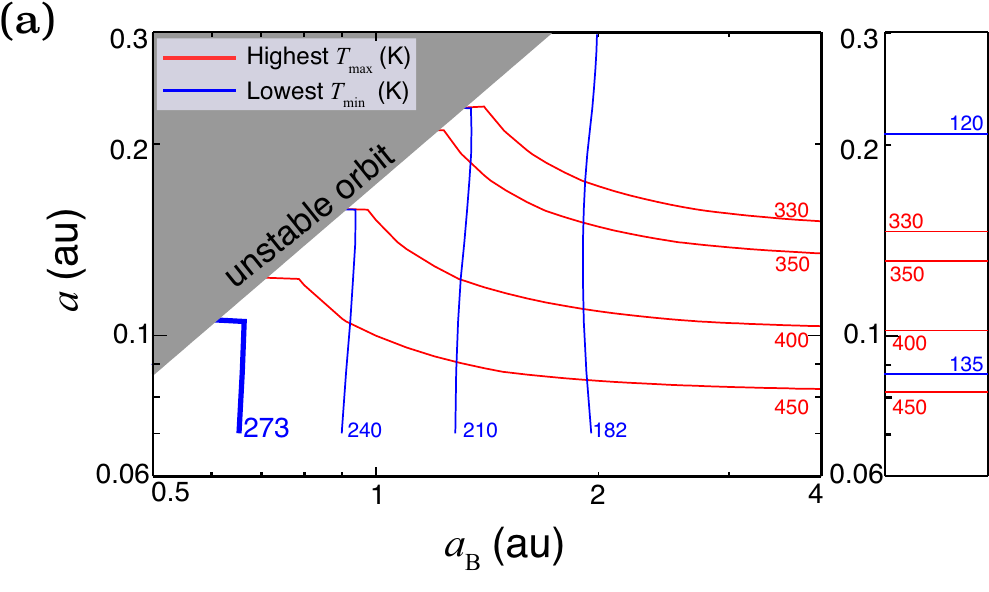}
\includegraphics[bb=0.000000 0.000000 285.380000 171.140000, width=0.51\textwidth]{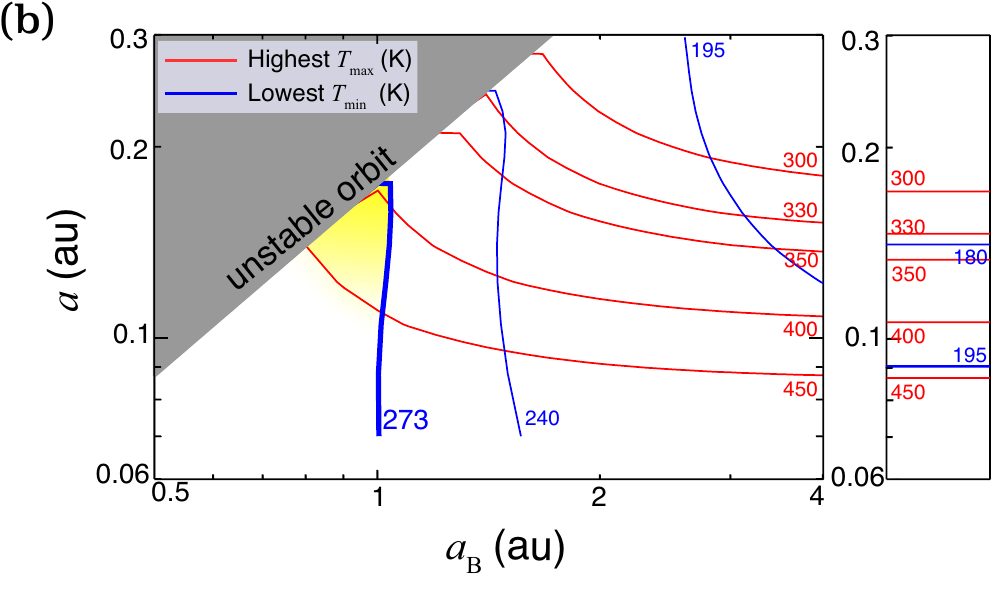}
\includegraphics[bb=0.000000 0.000000 284.900000 170.620000, width=0.51\textwidth]{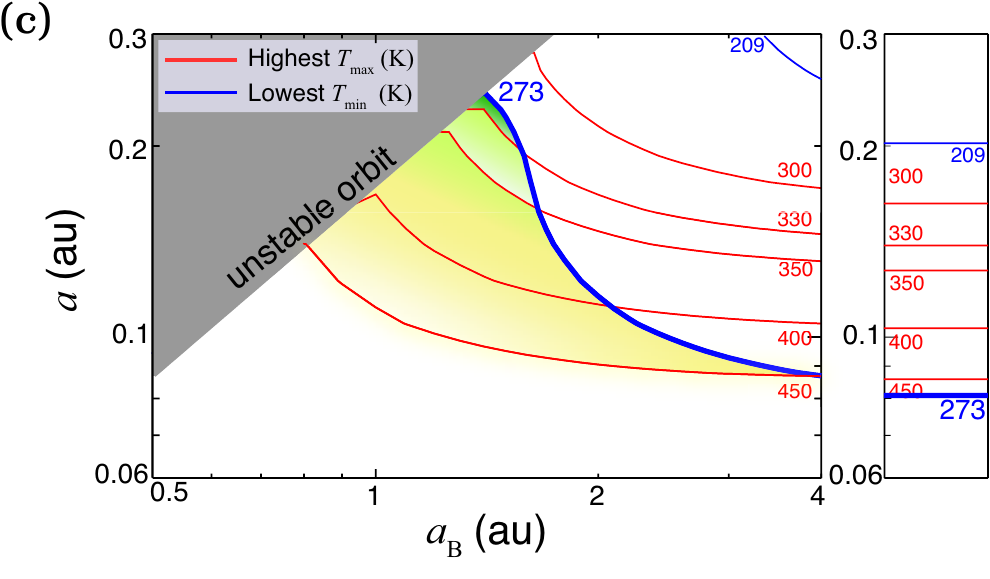}
\includegraphics[bb=0.000000 0.000000 287.000000 171.000000, width=0.51\textwidth]{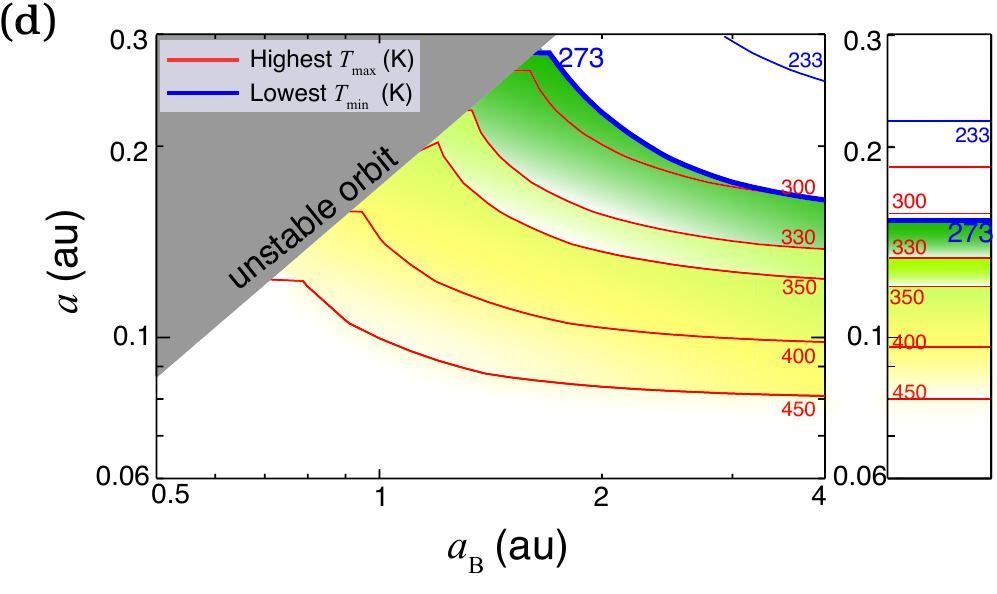}
\caption{Same as Figure~\ref{fig:BoundaryLand} except for the assumed atmosphere: 
CO$_{2}$ atmosphere of (a) 0.3, (b) 1, (c) 3, and (d) 10 bar.
}
\end{center}
\label{fig:BoundaryLandCO2}
\end{figure*}

Figure~\ref{fig:BoundaryLandCO2} shows the results for a CO$_{2}$ atmosphere.
Panel (b) shows the result of a CO$_{2}$ atmosphere of 1 bar.
Compared with the result of the Earth-like atmosphere (1 bar), the stronger greenhouse effect due to high CO$_{2}$ pressure increases temperature globally and shifts the \new{potentially habitable area}
to the larger $a$ and $a_{\rm B}$ area.
For $p{\rm CO}_2 < 0.3$ bar, the temperature difference between the dayside and the nightside
is so large that even G-type star radiation cannot produce the \new{potentially habitable area}.
For $p{\rm CO}_2 \sim 0.3$--3 bar, \new{the potentially habitable climate can be realized} 
only in the binary systems.
\new{For $p{\rm CO}_2 = 3$ bar, the G-type companion star at a distance of as far as $\sim4$ au can affect the potentially habitable areas.}
For $p{\rm CO}_2 > 3$ bar, while the efficient heat transport 
due to the thick atmosphere enables even the single M-type star case to have \new{area with habitable climate},
the \new{area} is much broader in the binary system case, especially for smaller $a_{\rm B}$ (except for
the case with too small $a_{\rm B}$ for the planetary orbital stability).
\new{For $p{\rm CO}_2=10$ bar, in the binary star case, the ratio $a_{\rm out}/a_{\rm in}$ of the area with Earth-like mild climate (green area) is at most twice of that in the case of the single M-type star.}

\subsubsection{\new{Temperature distribution for the constant total irradiance}}
\new{
Previous studies \citep{Kaltenegger+2013, Jaime+2014} estimated the HZs of S-type planets based on the total irradiance the planet receives from both stars and the orbital stability condition.
In this section, we fix the total irradiance and explore the impact of the horizontal distribution of the planetary surface, which the previous studies did not take into account.
}

\new{
In principle, for each value of binary separation $a_{\rm B}$, there is a
value of the semimajor axis of the planetary orbit $a$ for which the combined averaged incoming flux from the G-type and M-type stars is constant. 
We fix the averaged incoming stellar flux to $0.6S_{\odot}$, where $S_{\odot}$ is the solar irradiation flux at the substellar point, and calculate the 
surface temperature with changing $a_{\rm B}$. 
$0.6S_{\odot}$ is in the range of the stellar flux which a planet receives
in the classical HZ around a single M-type star and also in the HZ of a single G-type star \citep{Kopparapu+2013}.
The constant total irradiation line for 0.6 $S_{\odot}$ is plotted with black thick dashed lines in Figures~\ref{fig:BoundaryOcean} and \ref{fig:BoundaryLand}. 
}

\begin{figure*}
 \begin{minipage}{0.5\hsize}
  \begin{center}
   \includegraphics[bb= 0.000000 0.000000 385.850000 289.510000,width=1\textwidth]{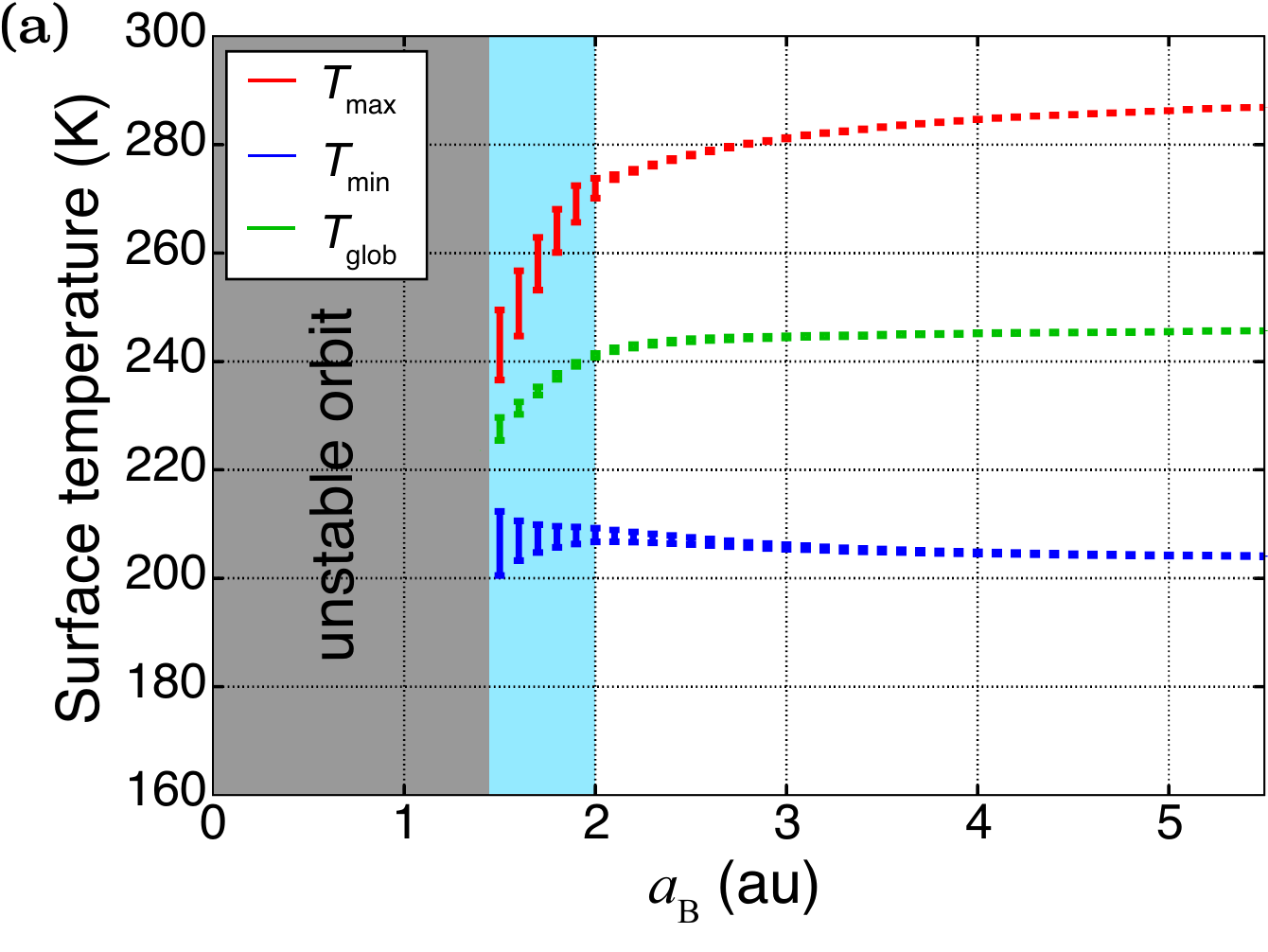}
  \end{center}
 \end{minipage}
 \begin{minipage}{0.5\hsize}
  \begin{center}
   \includegraphics[bb=0.000000 0.000000 385.900000 289.510000,width=1\textwidth]{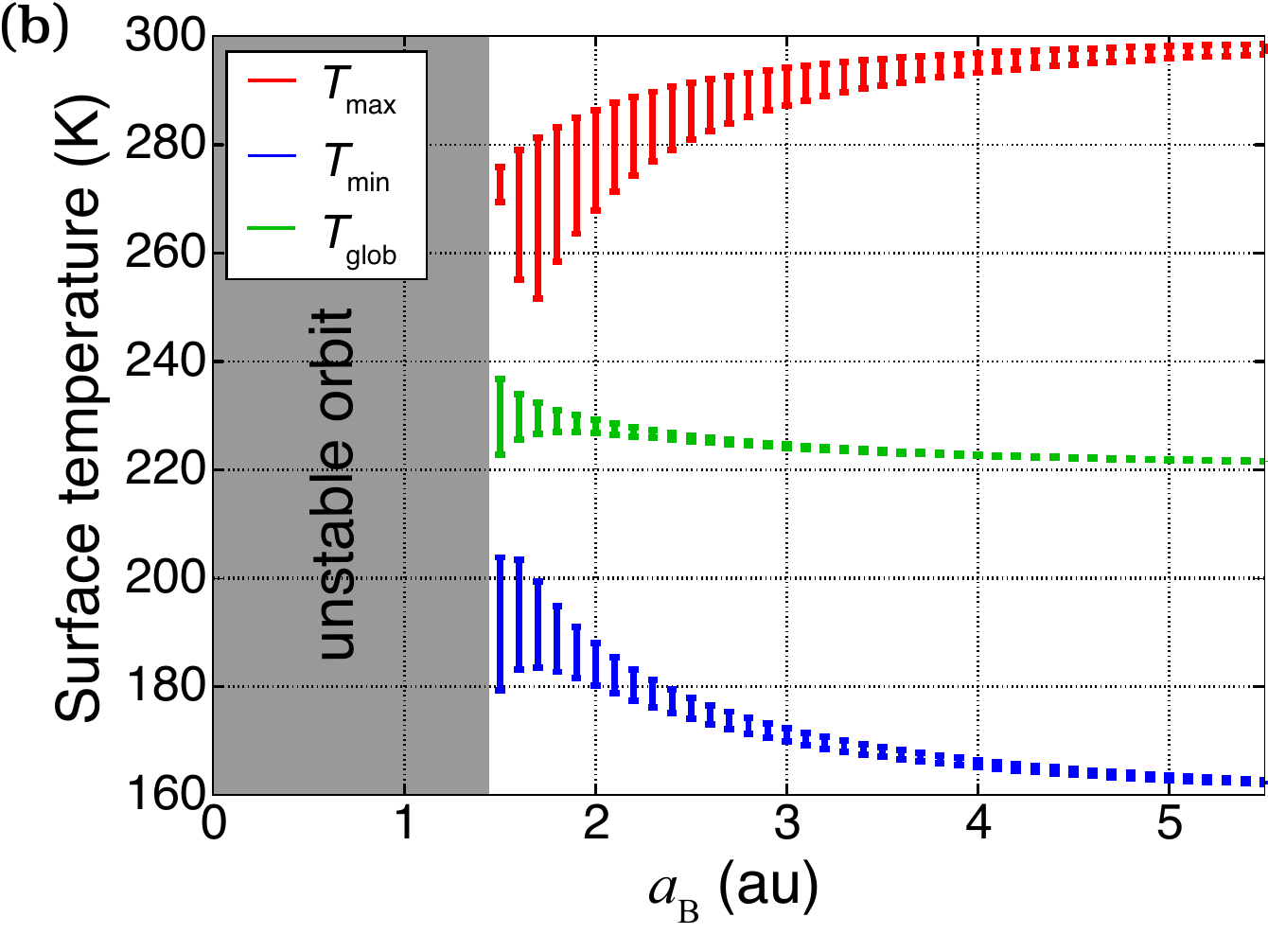}
  \end{center}
 \end{minipage}
\caption{The planetary surface temperature profile of (a) the ocean planets and (b) the land planets as 
a function of the binary separations $a_{\rm B}$ for a fixed total irradiance of $0.6S_{\odot}$. 
The red and blue plots are the maximum and minimum temperatures, respectively.
The green plots are the global-mean surface temperature.
The bars show the temporal variation in one synodic period. 
The cyan area reprensents the area where ocean globally freezes.
}
\label{fig:totalflux}
\end{figure*}

\newest{
Figure~\ref{fig:totalflux} shows the global-mean surface temperature ($T_{\rm glob}$), $T_{\rm max}$, and $T_{\rm min}$ as a function of $a_{\rm B}$ along this constant-total-irradiation line. 
$T_{\rm max}$ and $T_{\rm min}$ change with $a_{B}$ (and simultaneously $a$) as much as 50~K.
As a result, the planets would have diverse climates under the same total irradiation.
We highlight that the ocean planets undergo global glaciation when $a_{\rm B}\leq2.0$ au, the cyan region in the left panel of Figure~\ref{fig:totalflux}. 
In Figure~\ref{fig:BoundaryOcean} of ocean planets, we can see that indeed this region corresponds to the outside of the Earth-like climate area. 
This demonstrates that the total irradiance alone is not diagnostics for the habitable condition.
}
 
 \newest{
It is worth noting that $T_{\rm glob}$ increases with $a_{\rm B}$ for the ocean planets but decreases with $a_{\rm B}$ for the land planets.
These behaviors are explained by two factors.
First, because the assumed albedo depends on the stellar type, and the total {\it absorbed} flux, $F_{\rm M}(1-\alpha_{\rm M} ) + F_{\rm G}(1-\alpha_{\rm G} )$, varies with $a_{\rm B}$ (or $a$) even if $F_{\rm M}+F_{\rm G}$ = const. Because the albedo for the G-type star is larger in our model (Table \ref{tab:experiment}), this has the effect of increased total absorbed energy as $a_{\rm B}$ increases ($a$ decreases). This is qualitatively consistent with the behavior of ocean planets where $T_{\rm glob}$ increases as $a_{\rm B}$ increases. However, the behavior of land planets is opposite. This points us to the effect of the horizontal temperature distribution; Although the irradiation from the M-type star increases as $a_{\rm B}$ gets larger ($a$ gets smaller), the nightside warming is limited and is not sufficient to compensate the decrease of the irradiation from the G-type star. This leads to the decrease of $T_{\rm min}$ and $T_{\rm glob}$ as $a_{\rm B}$ increases. The balance between these two counteracting effects determines the climatological trend along the constant-total-irradiation line.
}

\subsubsection{Effect of binary eccentricity} \label{subsec:eccentricity}

So far, we have fixed the binary eccentricity, $e_{\rm{B}}$, to be 0. 
We also performed the run for a land planet with $e_{\rm{B}}=0.3$. The value of 0.3 is typical for binaries with periods from 10 days to 1000 days, which we investigate in this paper \citep{Duquennoy+1991}.

\begin{figure*}
\begin{center}
\includegraphics[bb=0.000000 0.000000 283.820000 163.410000, width=0.6\textwidth]{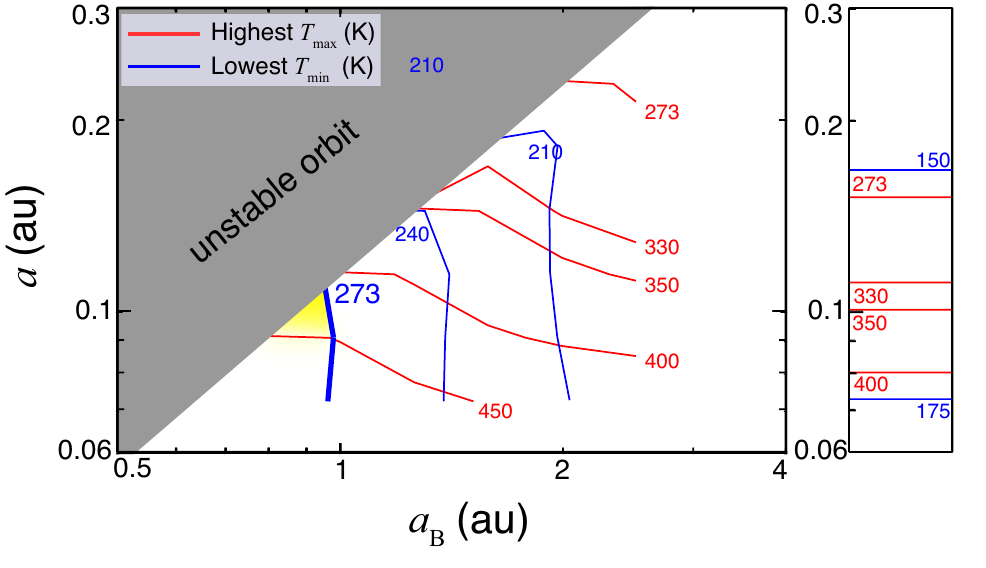}
\caption{Same as Figure~\ref{fig:BoundaryLand}
except for $e_{\rm B}=0.3$.}
\end{center}
\label{fig:e03}
\end{figure*}

Figure \ref{fig:e03} shows the $T_{\rm max}$ and $T_{\rm min}$ map and the \new{potentially habitable area} as a function of $a$
and $a_{\rm B}$ for land planets with Earth-like atmospheres in the $e_{\rm{B}}=0.3$ case.
Because $e_{\rm B} > 0$, we use $T_{\rm max}$ and $T_{\rm min}$, 
taking into account the change in the distance between the two stars.
Compared to the $e_{\rm B}=0$ case in Figure \ref{fig:BoundaryLand},
both the orbital stability limit and \new{potentially habitable area} are shifted to larger $a_{\rm B}$ for a fixed $a$.
The shift of the orbital stability limit is straightforwardly understood from the expression for the condition for stable orbits:
$a < a_{\rm max}\simeq 0.2 (1-e_{\rm{B}})^{1.2} a_{\rm{B}}$ (Eq.~\ref{eq:a_max}), 
or $a_{\rm B} >  a_{\rm B,min} \equiv 5 (1-e_{\rm{B}})^{-1.2} a $. 
Given $a$, the stability limit, $a_{\rm B,min}$, in the case of $e_{\rm B}=0.3$ is larger by a factor of $\sim 1.5$ than in the case of $e_{\rm B}=0$. 
On the other hand, the shift in $T_{\rm min}$ can be considered as follows:  
the highest value of the local $T_{\rm min}$ during the synodic period is controlled by the minimum distance between the G-type star and the planet, which is $(1-e_{\rm{B}})a_{\rm{B}} - a$. 
Thus, with a fixed $a$, the effect of the binary star with $e=0$ and $a_{\rm B}$ is approximately equivalent to that of a binary star with $e=0.3$ and $0.7^{-1} a_{\rm B}$. 
Because the cold-trap limit determined by $T_{\rm min}$=273K 
was $a_{\rm B} \sim 0.7$ au (Figure~\ref{fig:BoundaryLand}) in the case of $e_{\rm B}=0$, 
the limit is $a_{\rm B} \sim 1.0$ au in the case of $e_{\rm B}=0.3$ (Figure~\ref{fig:e03}).
Therefore, the \new{potentially habitable area} in an orbitally stable region in the $e_{\rm B}=0.3$ case
is shifted by 40--50\% in $a_{\rm B}$ from the $e_{\rm B}=0$ case.

\section{Discussion} \label{sec:discussion}
\subsection{The sensitivity to diffusion coefficient} \label{subsec:transport}

Our estimates of \new{the habitable orbital region} are based on the maximum and minimum surface temperature ($T_{\rm max}$ and $T_{\rm min}$), which are controlled by the diffusion coefficients, $D_{1}$ and $D_{2}$. 
While we determined their fiducial values referring to data for the Earth and GCM simulations for tidally locked planets, there are large uncertainties in their appropriate values, as they ultimately depend on the atmospheric and oceanic motions as well as latent heat transfer, which requires more complex modeling beyond the EBMs. 
\new{In addition,  it should be noted that for tidally locked planets, $D_{1}$ and $D_{2}$ depend on the planetary semimajor axis $a$ through the spin rotation rate which affect the atmospheric and oceanic motions.} 
Therefore, in this section, we discuss how the assumed values for the diffusion coefficients affect \new{our} predictions. \new{For convenience of the discussion, we adopt $T_{\rm max}$ equal to maximum temperature of present-day Earth or 450K as the inner limit of such area while the outer boundary is set for each type of planets based on section \ref{subsec:boundary}.}

In general, larger $D_{1,2}$ provides efficient heat re-distribution, decreasing $T_{\rm max}$ and increasing $T_{\rm min}$. 
For ocean planets where \new{the  boundary for global freezing is} mostly determined by $T_{\rm max}$, this means that the \new{the temperate climate zone} as a whole \new{should move} closer to the M-type star. 
In addition, higher $T_{\rm min}$ makes the planet less susceptible to atmospheric collapse, \new{which potentially extends the orbital region with temperate climate}
 toward larger $a_B$. 
For land planets, the smaller $T_{\rm max}$ moves the inner boundary of the \new{habitable area} inward, and the larger $T_{\rm min}$ tends to inhibit atmospheric collapse and/or cold trap.
Both effects broaden the \new{area where land planets could maintain habitable climate} with larger $D_{1,2}$.

\begin{figure}[htbp]
 \begin{minipage}{0.5\hsize}
  \begin{center}
   \includegraphics[bb= 0.000000 0.000000 379.700000 292.960000,width=1\textwidth]{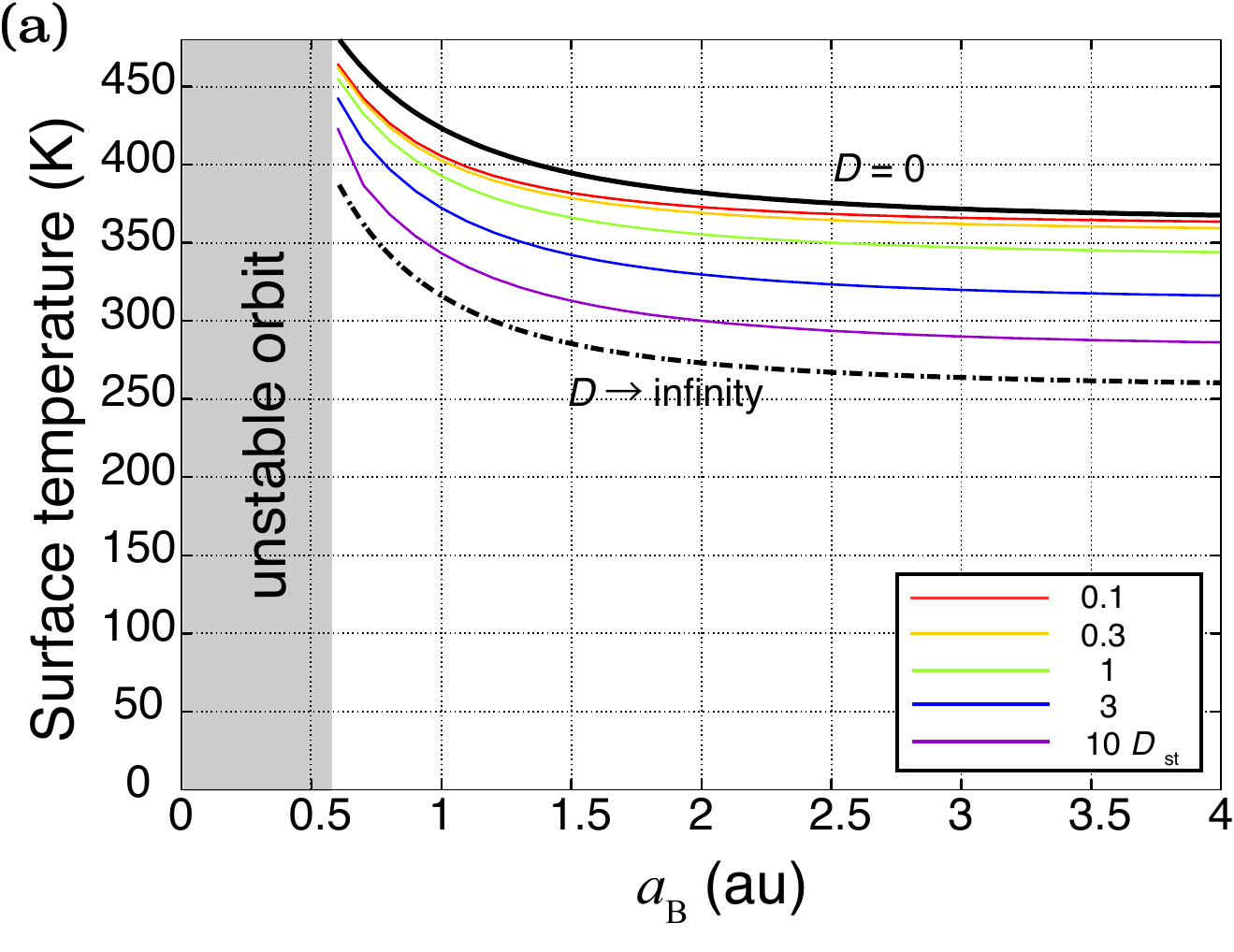}
  \end{center}
 \end{minipage}
 \begin{minipage}{0.5\hsize}
  \begin{center}
   \includegraphics[bb=0.000000 0.000000 379.700000 292.960000,width=1\textwidth]{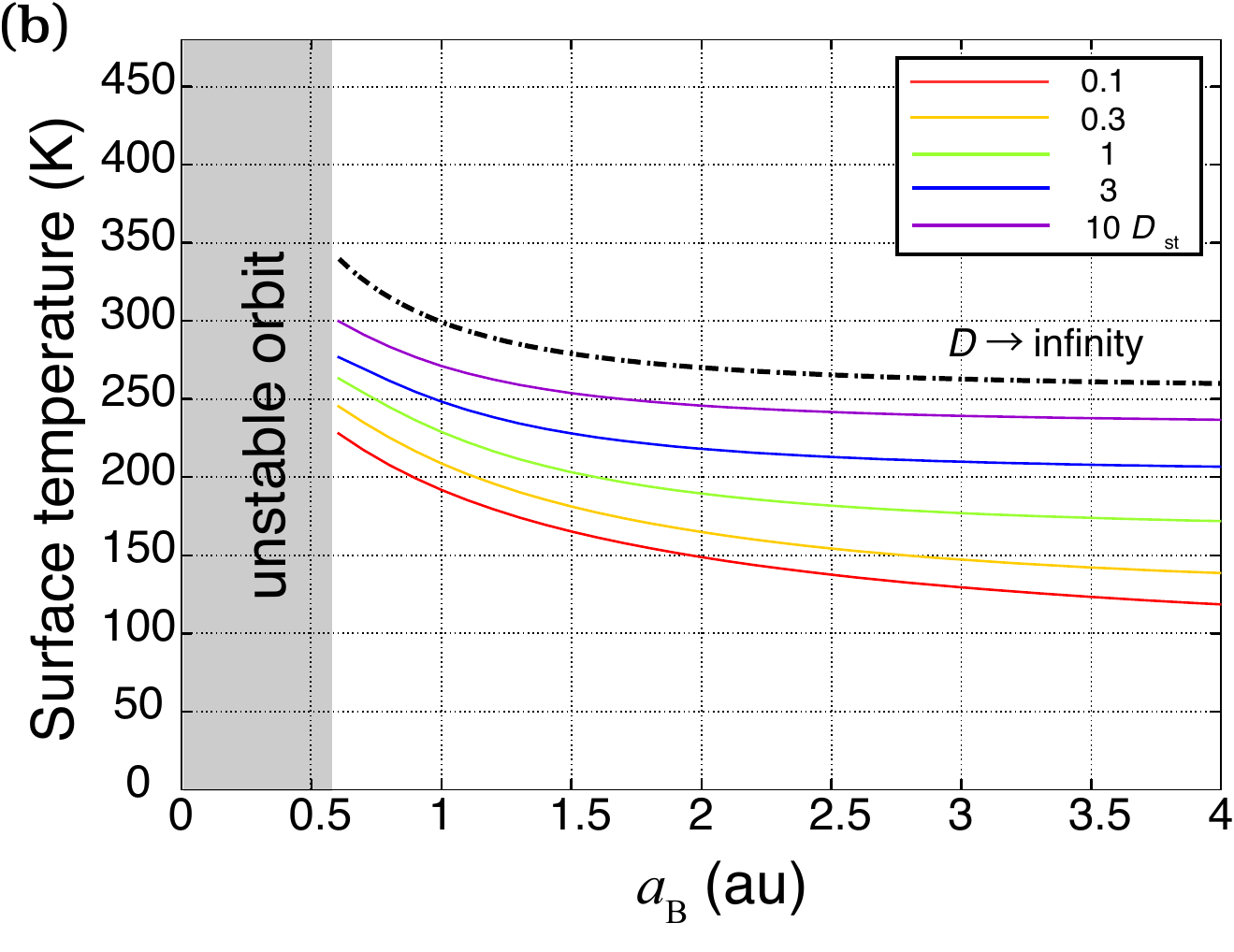}
  \end{center}
 \end{minipage}
 \caption{The planetary temperature profile of (a) the maximum temperature $T_{\rm max}$ and (b) minimum temperature $T_{\rm min}$ as a function of the binary separations $a_{\rm B}$ for a fixed planetary semimajor axis $a=0.1$ au.
The diffusion coefficients, $D_{1}$ and $D_{2}$, take 0.1, 0.3, 1, 3, 10 times as much as the values on a land planet with an Earth-like atmosphere. The black solid line is estimated for the case of $D=0$, and the dot-dash line is estimated for the case of $D\rightarrow\infty$. The shaded area is the orbitally unstable region (Eq. 1).
}
\label{fig:kakusanT}
\end{figure}

\begin{figure*}
\begin{center}
\includegraphics[bb=0.000000 0.000000 287.820000 163.410000, width=0.7\textwidth]{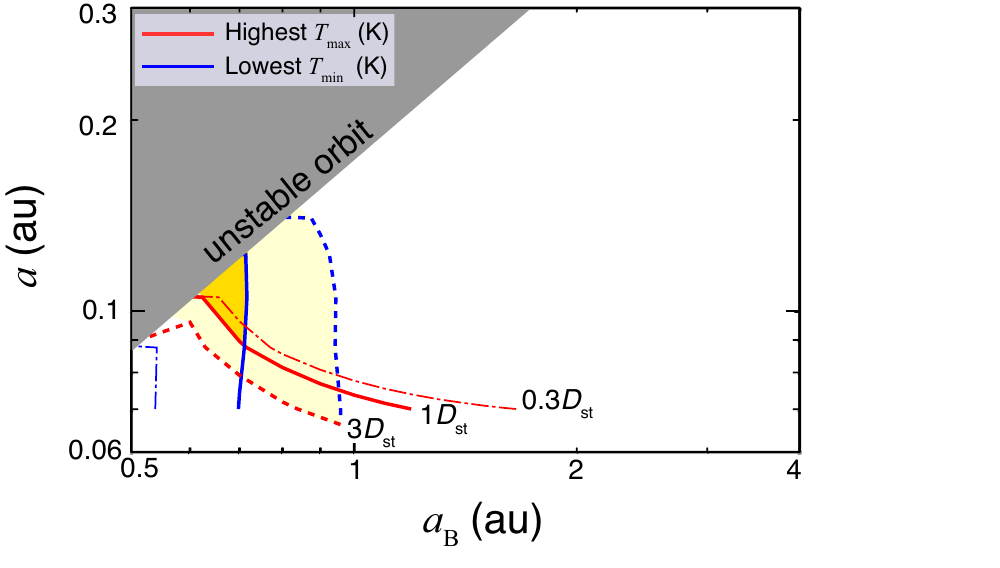}
\caption{Same as Figure~\ref{fig:BoundaryLand}
except for the diffusion coefficients; $D_{1,2} =0.3D_{\rm st}$ (dotted-dash lines),  $D_{1,2} =1D_{\rm st}$ (solid lines), and $D_{1,2} =3D_{\rm st}$ (dotted lines).
\new{The blue and red lines are the outer and inner boundaries for the potentially habitable area, corresponding to $T_{\rm min}=273$K and $T_{\rm max}=450$K, respectively.}}
\end{center}
\label{fig:kakusanHZ}
\end{figure*}

Here, we quantitatively demonstrate the change of the \new{orbital area with habitable climate} with varying diffusion coefficients, taking land planets as an example. 
We multiply the fiducial values of the diffusion coefficients $D_{\rm st}$ by 0.1--10, and repeat the same EBM simulations.
Figure \ref{fig:kakusanT} shows the resultant $T_{\rm max}$ (left panel) and $T_{\rm min}$ (right panel) as a function of the binary separation in the case of $a$ = 0.1 au. 
Two limits, the case of no heat transport ($ D_{1,2} = 0 $) and the case of globally uniform temperature ($ D_{1,2} \rightarrow \infty $) are also plotted for the result of $T_{\rm max}$.
The variation of $D_{1,2}$ from $0.1D_{\rm st}$ to $10D_{\rm st}$ changes $T_{\rm max}$ by $\sim$ 100 K, while beyond this range $T_{\rm max}$ asymptotically approaches the two limiting cases. 
Because the variation is larger for $T_{\rm min}$ than $T_{\rm max}$, the outer boundary of the \new{habitable climate area} is more sensitively affected by the values of $D_{1,2}$ than the inner boundary, as shown in Figure \ref{fig:kakusanHZ}.
If $D_{1,2}=3D_{\rm st}$, the outer boundary of $a_{\rm B}$ is shifted from 0.7 au to 1.0 au, doubling the range of the \new{potential habitable area} in the $a_{B}$ direction (Figure \ref{fig:kakusanHZ}).
If $D_{1,2}$ becomes sufficiently large, such an area appears even in the case of a single M-type star, although the width of the \new{area} is still larger in the case of the binary system, such as in the results in Figure \ref{fig:BoundaryLandCO2} (c) and (d). 
On the other hand, if $D_{1,2}=0.3D_{\rm st}$, the outer boundary is shifted to the small $a_{\rm B}$ area, overlapping with the orbitally unstable region.

\subsection{Beyond EBM} \label{subsec:beyond}

The core of our model in this paper is EBM. 
While EBM is useful to reveal the general trend of the planetary climate in question, this simplified approach inevitably has uncertainties in the parameters that cannot be determined within the framework of EBM. 
In this subsection, we mention the processes that are not explicitly captured in our EBM and discuss the uncertainties in our model. 

\paragraph{Diffusion coefficients and albedo} 

As stated in Section \ref{subsec:transport}, diffusion coefficients represent many processes in the planetary surface layers: atmospheric dynamics, oceanic flow, and the phase change and transport of water. 
These processes are also associated with cloud cover, which affects the planetary albedo. 
These processes depend on the irradiation pattern, spin rotation \new{and therefore the semimajor axis of the planetary orbit for tidally-locked planets,} atmospheric constituents, and surface pressure, among others. 
Therefore, adopting these values for planets in a broader parameter space is not trivial. 
Assessing the dependence of these processes  with full 3-D GCM simulations will be a future work.

\paragraph{Thermal emission as a function of temperature}

In our model, the relation between thermal emission and surface temperature is estimated from the 1-dimensional radiative-convective model. 
Here, it should be noted that thermal emission in principle depends not only on the atmospheric constituents, but also on the vertical thermal profile of the atmosphere. 
In our fiducial model, we assumed the equilibrium vertical temperature profiles obtained under the M-type star's spectrum and derived the fitting parameter, $\sigma'(p \rm{CO_{2}})$ in Eq. (\ref{eq:blarad-CO2}). 
When the irradiation from the G-type companion star becomes significant, however, the vertical temperature profile would be influenced by the spectrum of the G-type star. 
In the following, we evaluate the effect of such uncertainties in the vertical thermal profile of atmosphere, taking the land planet case as an example.

In order to estimate the range of uncertainties in the parameterization of planetary thermal emission, we performed additional radiative-convective calculations under the spectrum of the G-type star and derived $\sigma'(p \rm{CO_{2}})$ for that irradiation, which we call $\sigma_{\rm{G}}'(p)$. 
The difference between $\sigma_{\rm{G}}'(p)$  and the fiducial $\sigma '(p)$ are presented in Table \ref{tab:sigma}; it is shown that $\sigma_{\rm{M}}'(p)$ is larger than $\sigma_{\rm{G}}'(p)$ for all the atmospheric pressures, which is associated with a reduced vertical temperature gradient. 
The change in $\sigma '(p)$ would be accounted for by the change in surface temperature,
\begin{equation}
T'_{\rm G} =(\sigma_{\rm{M}}'(p)/\sigma_{\rm{G}}'(p))^{1/4}\times T'_{\rm M}.
\end{equation}
Because even the largest $(\sigma_{\rm{M}}'(p)/\sigma_{\rm{G}}'(p))^{1/4}$ is 1.06 with 10 bar atmosphere, the vertical profile
 of the atmosphere under the G-type star's radiation would increase the temperature   by only $\sim$ 15K \new{at the boundary for the cold trap of water} \new{i.e.,} $T_{\rm min}=273$K.
 The real solution under the irradiation from both an M-type star and G-type star would be somewhere in between. 
 Thus, the dependence on the vertical profile is not likely to change the global picture of our results.  

\begin{table*}[h!]
\tablenum{2}
\renewcommand{\thetable}{\arabic{table}}
\begin{center}
\caption{Values of blackbody radiation coefficient under each incoming wavelength \label{tab:sigma}}
\begin{tabular}{c|cccc}
\tablewidth{0pt}
\hline 
\hline 
CO$_{2}$ pressure & 0.3bar &1bar & 3bar & 10bar \\
\hline
$\sigma_{\rm{G}}'(p)$\ [W m$^{-2}$ K$^{-4}]$ & $3.88\times 10^{-8}$ & $2.90\times 10^{-8}$ & $2.09\times 10^{-8}$ & $1.42 \times 10^{-8}$ \\
$\sigma_{\rm{M}}'(p)$\ [W m$^{-2}$ K$^{-4}]$ & $4.12\times 10^{-8}$ & $3.19\times 10^{-8}$ & $2.37 \times 10^{-8}$ &$1.78\times 10^{-8}$ \\
\hline
\end{tabular}
\tablecomments{The fitting parameters in Eq. (\ref{eq:blarad-CO2}) obtained under the vertical profile of the atmosphere with the G-type star's radiation and the M-type star's radiation ($\sigma_{\rm{G}}'$ and $\sigma_{\rm{M}}'$) (see details in Appendix \ref{ap:radconv}).
}
\end{center}
\end{table*}

\new{
\subsection{HZ of tidally locked planets \label{subsec:tidally-HZ}}
}
\new{
In this paper, we estimate the planetary climate based on the maximum and minimum temperatures, and the threshold value for the inner limit is adopted in reference to the maximum surface temperature of present-day Earth.
On the other hand, the inner limit of conventional habitable zones  \citep{Kasting+1993} is determined by runaway greenhouse effect, and EBM by itself is not able to produce such a climate instability.
According to previous works on the tidally locked planet around a single M-type
star, the inner boundary condition of the HZ would be set as follows.
}

\new{
Ocean planets have an upper limit for outgoing flux \citep[e.g.,][]{Ingersoll1969,Nakajima+1992,Kasting+1993}. 
If a planet receives insolation beyond the limit, the planet undergoes the runaway greenhouse effect and it sets the inner boundary.
Tidally locked ocean planets are relatively stable against the high irradiance because of the high albedo due to cloud decks \citep{Yang+2013, Kopparapu+2016}. Under sufficiently strong irradiation, however, the increasing atmospheric water vapor heats the atmosphere, which eventually forms a temperature inversion, suppresses the convection, and dismisses the cloud decks in the dayside \citep{Kopparapu+2017}. As a result, the planet transits into the runaway greenhouse state. \citet{Kopparapu+2017} showed that this transition for a tidally locked ocean planet around a single M3-type star occurs when the maximum surface temperature exceeds 300-310 K.
}

\new{
On the other hand, on a land planet, water tends to be transported to the cooler nightside, and the dayside becomes dry, from which more infrared radiation is emitted. 
\citet{Abe+2011}, a GCM study for rapidly rotating land planets, showed that the boundary of the runaway greenhouse effect moves inward to the central star. 
This is because the runaway greenhouse initiates when the water-trapping regions rather than the maximum-temperature region become warm enough \citep{Kodama+2018}, and the same trend would be applied to a tidally locked land planet.
As a result, the runaway greenhouse state of land planets may correspond to higher temperature than on ocean planets, 400-500 K (T. Kodama 2018, private communication). Based on these trends, we demonstrate the potentially habitable areas (yellow areas) of the land planets as well as areas (green areas) with Earth-like climate in Figures~\ref{fig:BoundaryLand}, \ref{fig:BoundaryLandCO2}, and \ref{fig:e03}.
}

\new{
However, the exact conditions for tidally locked planets to undergo the runaway greenhouse regime is not fully understood even around single stars, particularly for tidally locked planets. The criteria for the boundary of the runaway greenhouse involves various atmospheric processes and the stellar spectral type. The climate may also be history-dependent \citep{Leconte+2013}.
As a future work, it is important to examine the climates of tidally locked land planets around the inner edge of the HZs.
}

\subsection{Observability} \label{subsec:obs}
In this section, we discuss future possibilities of detecting and characterizing such planets as were studied in this paper; specifically, an S-type Earth-sized planet around an M-type star that has a G-type companion star at a distance of a few au. 

As for detection, the standard methods that have been most successful for discovering planets around single stars (or those in binary systems with large separation) would be challenged by the presence of the G-type companion stars. 
For example, radial velocity measurements will be more complex due to the Doppler motion determined by the three-body problem, likely calling for further development in data analysis techniques. 
In transit surveys, the (spatially unresolved) G-type companion star, which is $\sim 10^2$ times brighter than the M-type star, decreases the transit depth by the order of 10$^{-2}$. Therefore, the transit signal of an Earth-sized planet in front of an M-type star, order of $10^{-3}$ relative to the flux of the M-type star, would be as small as $\sim 10^{-5}$ of the total flux, below the detection limit $\sim 10^{-4}$. Larger terrestrial planets could be within reach. 

Recently, \citet{Oshagh+2017} proposed a method for S-type planets in eclipsing binaries, which uses correlation between the RVs, eclipse timing variations (ETVs), and eclipse duration variations (EDVs). Future missions, such as PLATO (scheduled for launch in 2025), will give the targets of eclipsing binaries to which this method may be applied. 

Once they are discovered, they will be interesting targets to follow up with detailed observations to characterize their atmospheres. 
Direct imaging of such planets requires future coronagraphic facilities that can suppress starlight to the order of $10^{-10}$ (the contrast between the planet and the G-type star in the visible or near-infrared) while having a small inner working angle to resolve the planet and the M-type star (a few tens of milliarcseconds). Such a facility is beyond the capability of the planned projects \citep[see Figure 7 of ][]{Fujii+2018}. 
Transit transmission spectroscopy is probably more promising. 
Assuming a 30-meter-class telescope, the diffraction-limited point spread function 
at 30 mas is 10$^{-3}$, hence the contamination from the G-type star at a distance of 1 au from the M-type star 30 pc away is $10^{-1}$. This could be further suppressed when coronagraphic instruments are used, potentially to the level of no interference with planetary signals. 
If the G-type star cannot be suppressed enough to be negligible, any variations from the G-type star must be carefully removed. This may be possible through high-resolution spectroscopy, using their different Doppler shifts in lines: the M-type star and the G-type star orbits around their barycenter, whose radial velocity is approximately $>$ 3 km/s in amplitude and has opposite phase. As the radial velocity of the planet is similar to that of the M-type star, it may not be difficult to separate the spectral lines due to planetary atmospheres from the spectral features in G-type star spectra.

\section{Conclusion} \label{sec:conclusion}
The planets in the habitable zones (HZs) around M-type stars are likely to be tidally locked because the HZs are close to the central star \citep{Kasting+1993}.
The planets with thin ($<$ a few 0.1 bar) atmospheres suffer from an extreme temperature difference between the dayside and nightside, which could lead to the condensation of localized water (cold trap) and atmospheric collapse on the cold nightside.
This is one of the crucial problems for the habitability of a planet around a single M-type star \citep{Joshi+1997, Leconte+2013, Turbet+2016, Turbet+2017}. 

If the planets have thick atmospheres, this problem could be solved by efficient atmospheric heat transport \citep[e.g.][]{Joshi+1997}.
We have found that, for the planets with thin atmospheres, the problem can be solved
if the planet-hosting M-type star has a much brighter binary companion star such as a G-type star.
While the mass difference between G-type stars and M-type stars is not so large, the luminosity of G-type stars are a few orders of magnitude brighter. This enables a G-type star to warm up the cold nightside of the planet around the M-type star without destabilizing the planetary orbit, and avoids cold trap and atmospheric collapse.

We investigated the \new{surface} temperature distribution of the tidally locked planets
irradiated by an M-type host star and a G-type companion star 
through simulations of the two-dimensional energy balance model (EBM)  
\citep[e.g.][]{North1975}, which is calibrated by global circulation model (GCM) results.
Because EBM is simpler and computationally less expensive than GCM,
we surveyed a broad range of parameters: the planetary orbital radius, binary separation, planetary surface (land-covered with limited surface water or ocean-covered), and atmospheric compositions/pressure.

In this paper, we focus on the habitable climate which enables planets to have liquid water on their surface.
We found the following:
\begin{enumerate}
\item The irradiation from the G-type star is much more effective
on the nightside of the planet than on the dayside for two reasons: 
the temperature contributed by the M-type host star is very low on the nightside, and
the distance between the planet and the G-type star is the smallest when the star irradiates the nightside of the planet. 
\new{This} effect is more pronounced for land planets.

\item 
\newest{
Although ocean planets around a single M-type star 
do not become habitable with CO$_2$ atmospheric pressure $p \la 0.3$ bar due to atmospheric collapse, the G-type star's irradiation within $a_{\rm B}$ of $\sim 2.5$ au helps to provide a temperate climate.}

\item 
\newest{While the land planets around a single M-type star 
do not produce a moderate Earth-like climate for an atmospheric pressure of $p \la 10$ bar due to the cold trap, those around M-type stars with a G-type companion star are able to have such a climate if $p \ga 3$ bar and if the binary separation is $a_{\rm B} \sim 1-2$ au. 
If land planets can have stable liquid water far beyond the Earth-like temperature regime, as suggested by some GCM studies \citep[e.g.,][]{Abe+2011,Kodama+2018}, planets with thinner atmospheres ($\ga 0.3$~bar) can also be habitable when orbiting a star that has a companion star.}

\item
\newest{
Even if the total irradiation is the same, the climates of tidally locked S-type planets vary from Earth-like temperate climates to completely frozen ones, depending on the orbital configuration of the system. 
}

\end{enumerate}

We also performed runs of the land planets for the binary eccentricity $e_{\rm B}=0.3$
to find that the \new{potentially habitable areas} are only shifted to the larger binary separation area without any qualitative change.
The \new{planetary climate and habitable conditions}
 should depend on other factors, such as 
the atmospheric/oceanic dynamics, water distribution, clouds, and \new{planetary spin rotations}.
These issues must be studied in detail by full 3-dimensional GCM simulations. 
On the other hand, EBM has the advantage of much lower computational costs,
which enables us to understand intrinsic physics by 
exploring broad parameter space. 
The combination of GCM and EBM would be important to
clarify the climate of exoplanets, especially those having complicated 
configurations such as the system we studied in this paper.

Although it is not easy at present to detect S-type planets in close
binary systems, S-type planets have been discovered in binary systems with the separation 
down to 5.3 au.
Future missions, such as PLATO, will be able to detect the S-type planets using new methods combining the RVs, ETVs, and EDVs \citep{Oshagh+2017}.
The 30-40 meter class future telescopes, GMT, TMT and the E-ELT, may also enable us to characterize such planets by the transmission spectroscopy of the atmosphere. About half of all G-type stars are known to have companion stars.
 According to the statistics, the binary systems comprised of an M-type star and a G-type star exist in a considerable number \citep{Duquennoy+1991, Raghavan+2010}. 
As we have showed here, the tidally locked planets around M-type stars with more luminous binary companion stars should be very interesting targets in terms of habitability.

\acknowledgments
\new{The authors thank the referee for constructive comments.}
We thank Teruyuki Hirano, Takanori Kodama, and Masahiro Ikoma for helpful comments and discussions.
We also thank David S. Amundsen for developing the k-coefficient tables for SOCRATES for CO$_2$ dominated atmospheres, which were used in the radiative-convective model as described in Appendix \ref{ap:radconv}.
This paper was supported by JSPS KAKENHI grant 15H02065 and 18H05438.

\vspace{5mm}
\software{SOCRATES \citep{Edwards_Slingo_1996, Edwards_1996}}

\newpage

\appendix

\section{Determining thermal flux of land planets with pure CO$_2$ atmospheres for a given surface temperature}
\label{ap:radconv}

In this section, we describe how we obtained the relation between the thermal radiation and the surface temperature for a land (dry) planet with a pure CO$_2$ atmosphere (equation (\ref{eq:blarad-CO2}), to be used in equation (\ref{eq:EBeq})).

In order to find a reasonable relation between these two factors that are independent of other parameters, we employed a 1-dimensional radiative-convective model. 
We developed a code to compute vertical temperature profiles in radiative equilibrium with the convective adjustment using the time-stepping method, following \citet{Manabe_Wetherald_1967}. 
For radiative transfer calculation in our model, we adopted the Suite Of Community RAdiative Transfer codes based on Edwards and Slingo \citep[SOCRATES;][]{Edwards_Slingo_1996, Edwards_1996}. 
SOCRATES was developed at the UK Met Office, and has been widely used for climate modeling \citep[e.g.,][]{Amundsen_et_al_2016, Way_et_al_2017}. 
SOCRATES uses the two-stream approximation for both long-wave (thermal) and short-wave (stellar) radiation, and opacities are treated using the correlated-k method \citep{Lacis_Oinas_1991, Goody_et_al_1989}, with k-terms for multiple gases combined using adaptive equivalent extinction \citep{Edwards_1996, Amundsen_et_al_2017}, see \citet{Way_et_al_2017} and \citet{Fujii_et_al_2017} and references therein for more details.
The opacities included in our calculation are: CO$_2$ self-broadening based on HITRAN2012 \citep{Rothman_et_al_2013}, CO$_2$ sub-Lorentzian line wings \citep{Perrin_Hartmann_1989, Wordsworth_et_al_2010}, and and CO$_2$-CO$_2$ collision induced absorption \citep{Gruszka_Borysow_1998, Baranov_et_al_2004, Wordsworth_et_al_2010}. 
In order to secure the accuracy, 17 bands and 42 bands are used for long-wave and for short-wave, respectively \citep[see Table 3 of][]{DelGenio_et_al_2018}. 
The Rayleigh scattering coefficient for CO$_2$ is calculated based on \citet{Bideau-Mehu_et_al_1973}. 

For the simulation of Figure \ref{fig:Proxima}, the spectrum of Proxima Centauri b ($T_{\rm eff}=3042$ K) taken from the Virtual Planetary Laboratory Molecular Spectroscopic Database \citep{Meadows_et_al_2016} was used to compute the temperature profile in the radiative-convective equilibrium. 
For other simulations, a modeled spectrum of a star with effective temperature of 3300 K, mass of 0.25 $M_{\odot }$, log $g$ of 5 (corresponding to radius of $0.3 R_{\odot }$), and zero metallicity was taken from the BT-Settl model \citep{Allard_et_al_2012} and used to find the radiative-convective equilibrium. 
The incident angle was set to $60^{\circ }$ and the additional factor of 0.5 was applied in order to match the globally averaged flux. 
The surface albedo was set at 0.2 (see Section \ref{subsec:EBM}). 

\begin{figure}[h]
\begin{center}
   \includegraphics[bb=0.000000 0.000000 360.000000 252.000000,width=0.6\hsize]{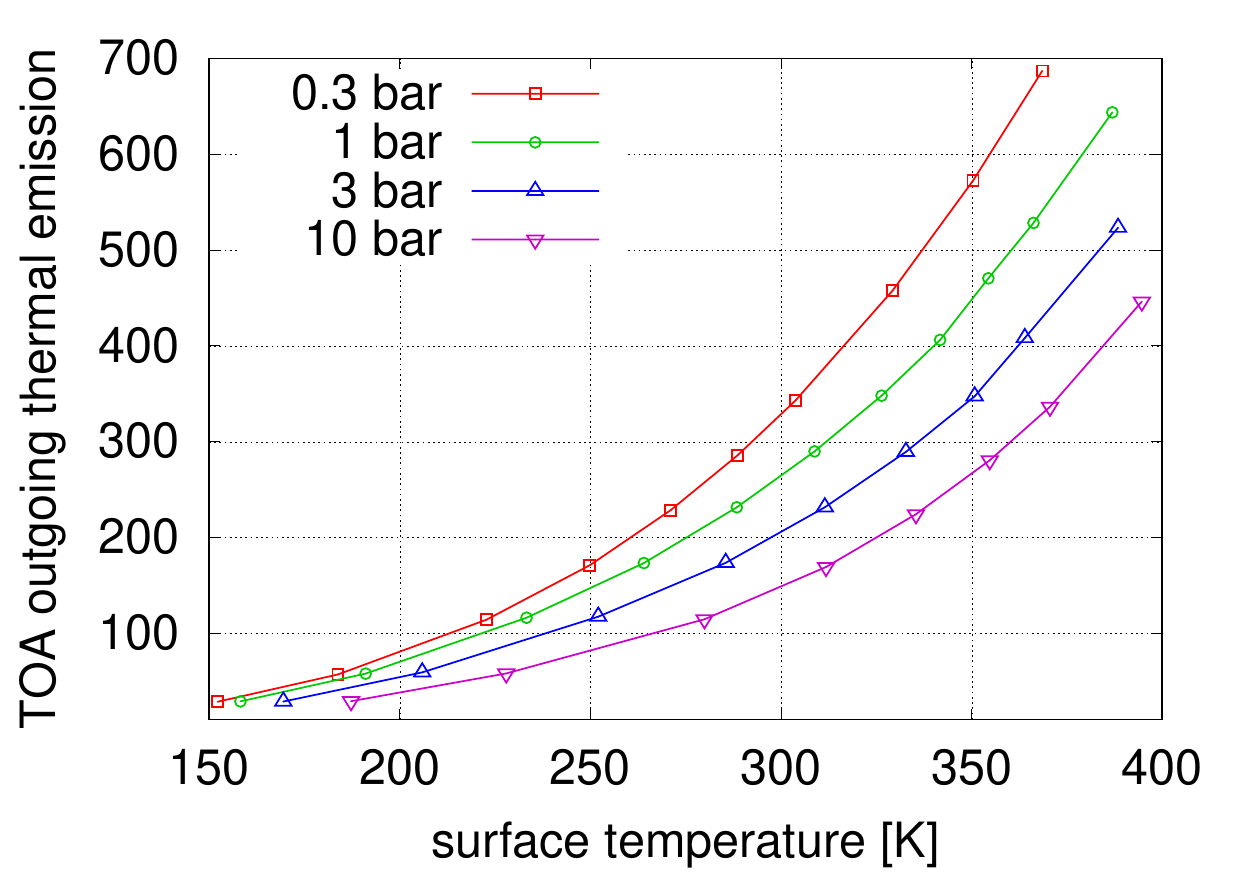}
    \caption{The relation between the top-of-atmosphere thermal flux and the surface temperature for a pure CO$_2$ atmosphere with varying surface pressures, based on 1-dimensional radiative-convective calculations. The incident spectra is modeled spectra of an M-type star with effective temperature of 3300 K, mass of 0.25 $M_{\odot }$, log $g$ of 5, and zero metallicity, with the incident angle of 60 degrees. Surface albedo is set at 0.2, and the surface gravity is the same as the Earth. }
    \label{fig:Fthermal_Tsurf}
  \end{center}
\end{figure}

For a given surface pressure of CO$_2$ (0.3, 1, 3 or 10 bar), the equilibrium vertical temperature profiles were computed for varying total incident flux, and the corresponding surface temperature, outgoing top-of-atmosphere long-wave (thermal) flux, and top-of-atmosphere short-wave albedo were recorded. 
The result are shown in Figure \ref{fig:Fthermal_Tsurf}. 
By fitting \new{the values within the range of TOA thermal flux between 150 W/m$^2$ and 350 W/m$^2$}, 
we obtained the approximate representations of outgoing thermal flux as a function of surface temperature described in Section \ref{subsec:EBM}.

\newpage
\begin{thebibliography}{}
\expandafter\ifx\csname natexlab\endcsname\relax\def\natexlab#1{#1}\fi

\bibitem[{{Abe} \& {Abe-Ouchi}(2005)}]{Abe+2005}
{Abe}, Y., \& {Abe-Ouchi}, A. 2005, AGU Fall Meeting Abstracts, P51D

\bibitem[{{Abe} {et~al.}(2011){Abe}, {Abe-Ouchi}, {Sleep}, \&
  {Zahnle}}]{Abe+2011}
{Abe}, Y., {Abe-Ouchi}, A., {Sleep}, N.~H., \& {Zahnle}, K.~J. 2011,
  Astrobiology, 11, 443

\bibitem[{{Allard} {et~al.}(2012){Allard}, {Homeier}, \&
  {Freytag}}]{Allard_et_al_2012}
{Allard}, F., {Homeier}, D., \& {Freytag}, B. 2012, Philosophical Transactions
  of the Royal Society of London Series A, 370, 2765

\bibitem[{{Amundsen} {et~al.}(2017){Amundsen}, {Tremblin}, {Manners},
  {Baraffe}, \& {Mayne}}]{Amundsen_et_al_2017}
{Amundsen}, D.~S., {Tremblin}, P., {Manners}, J., {Baraffe}, I., \& {Mayne},
  N.~J. 2017, \aap, 598, A97

\bibitem[{{Amundsen} {et~al.}(2016){Amundsen}, {Mayne}, {Baraffe}, {Manners},
  {Tremblin}, {Drummond}, {Smith}, {Acreman}, \&
  {Homeier}}]{Amundsen_et_al_2016}
{Amundsen}, D.~S., {Mayne}, N.~J., {Baraffe}, I., {et~al.} 2016, \aap, 595, A36

\bibitem[{{Artymowicz} \& {Lubow}(1994)}]{Artymowicz+Lubow1994}
{Artymowicz}, P., \& {Lubow}, S.~H. 1994, \apj, 421, 651

\bibitem[{{Baranov} {et~al.}(2004){Baranov}, {Lafferty}, \&
  {Fraser}}]{Baranov_et_al_2004}
{Baranov}, Y.~I., {Lafferty}, W.~J., \& {Fraser}, G.~T. 2004, Journal of
  Molecular Spectroscopy, 228, 432

\bibitem[{{Bideau-Mehu} {et~al.}(1973){Bideau-Mehu}, {Guern}, {Abjean}, \&
  {Johannin-Gilles}}]{Bideau-Mehu_et_al_1973}
{Bideau-Mehu}, A., {Guern}, Y., {Abjean}, R., \& {Johannin-Gilles}, A. 1973,
  Optics Communications, 9, 432

\bibitem[{{Boyajian} {et~al.}(2012){Boyajian}, {von Braun}, {van Belle},
  {McAlister}, {ten Brummelaar}, {Kane}, {Muirhead}, {Jones}, {White},
  {Schaefer}, {Ciardi}, {Henry}, {L{\'o}pez-Morales}, {Ridgway}, {Gies}, {Jao},
  {Rojas-Ayala}, {Parks}, {Sturmann}, {Sturmann}, {Turner}, {Farrington},
  {Goldfinger}, \& {Berger}}]{Boyajian+2012}
{Boyajian}, T.~S., {von Braun}, K., {van Belle}, G., {et~al.} 2012, \apj, 757,
  112

\bibitem[{{Caldeira} \& {Kasting}(1992)}]{Caldeira+1992}
{Caldeira}, K., \& {Kasting}, J.~F. 1992, \nat, 359, 226

\bibitem[{{Checlair} {et~al.}(2017){Checlair}, {Menou}, \&
  {Abbot}}]{Checlair+2017}
{Checlair}, J., {Menou}, K., \& {Abbot}, D.~S. 2017, \apj, 845, 132

\bibitem[{{Del Genio} {et~al.}(2018){Del Genio}, {Way}, {Amundsen}, {Aleinov},
  {Kelley}, {Kiang}, \& {Clune}}]{DelGenio_et_al_2018}
{Del Genio}, A.~D., {Way}, M.~J., {Amundsen}, D.~S., {et~al.} 2018, accepted by
  Astrobiology, arXiv:1709.02051

\bibitem[{{Dupuy} {et~al.}(2016){Dupuy}, {Kratter}, {Kraus}, {Isaacson},
  {Mann}, {Ireland}, {Howard}, \& {Huber}}]{Dupuy+2016}
{Dupuy}, T.~J., {Kratter}, K.~M., {Kraus}, A.~L., {et~al.} 2016, \apj, 817, 80

\bibitem[{{Duquennoy} \& {Mayor}(1991)}]{Duquennoy+1991}
{Duquennoy}, A., \& {Mayor}, M. 1991, \aap, 248, 485

\bibitem[{{Edwards}(1996)}]{Edwards_1996}
{Edwards}, J.~M. 1996, Journal of Atmospheric Sciences, 53, 1921

\bibitem[{{Edwards} \& {Slingo}(1996)}]{Edwards_Slingo_1996}
{Edwards}, J.~M., \& {Slingo}, A. 1996, Quarterly Journal of the Royal
  Meteorological Society, 122, 689

\bibitem[{{Fujii} {et~al.}(2017){Fujii}, {Del Genio}, \&
  {Amundsen}}]{Fujii_et_al_2017}
{Fujii}, Y., {Del Genio}, A.~D., \& {Amundsen}, D.~S. 2017, \apj, 848, 100

\bibitem[{{Fujii} {et~al.}(2010){Fujii}, {Kawahara}, {Suto}, {Taruya},
  {Fukuda}, {Nakajima}, \& {Turner}}]{Fujii+2010}
{Fujii}, Y., {Kawahara}, H., {Suto}, Y., {et~al.} 2010, \apj, 715, 866

\bibitem[{{Fujii} {et~al.}(2018){Fujii}, {Angerhausen}, {Deitrick},
  {Domagal-Goldman}, {Grenfell}, {Hori}, {Kane}, {Pall{\'e}}, {Rauer},
  {Siegler}, {Stapelfeldt}, \& {Stevenson}}]{Fujii+2018}
{Fujii}, Y., {Angerhausen}, D., {Deitrick}, R., {et~al.} 2018, Astrobiology,
  18, 739

\bibitem[{{Gong} \& {Ji}(2018)}]{Gong+2018}
{Gong}, Y.-X., \& {Ji}, J. 2018, \mnras, arXiv:1805.05868

\bibitem[{{Goody} {et~al.}(1989){Goody}, {West}, {Chen}, \&
  {Crisp}}]{Goody_et_al_1989}
{Goody}, R., {West}, R., {Chen}, L., \& {Crisp}, D. 1989, \jqsrt, 42, 539

\bibitem[{{Gruszka} \& {Borysow}(1998)}]{Gruszka_Borysow_1998}
{Gruszka}, M., \& {Borysow}, A. 1998, Molecular Physics, 93, 1007

\bibitem[{{Hartmann} \& {Larson}(2002)}]{Hartmann+2002}
{Hartmann}, D.~L., \& {Larson}, K. 2002, \grl, 29, 1951

\bibitem[{{Ingersoll}(1969)}]{Ingersoll1969}
{Ingersoll}, A.~P. 1969, Journal of Atmospheric Sciences, 26, 1191

\bibitem[{{Jaime} {et~al.}(2014){Jaime}, {Aguilar}, \& {Pichardo}}]{Jaime+2014}
{Jaime}, L.~G., {Aguilar}, L., \& {Pichardo}, B. 2014, \mnras, 443, 260

\bibitem[{{James} \& {North}(1982)}]{James+1982}
{James}, P.~B., \& {North}, G.~R. 1982, \jgr, 87, 10271

\bibitem[{{Joshi} {et~al.}(1997){Joshi}, {Haberle}, \& {Reynolds}}]{Joshi+1997}
{Joshi}, M.~M., {Haberle}, R.~M., \& {Reynolds}, R.~T. 1997, \icarus, 129, 450

\bibitem[{{Kaltenegger} \& {Haghighipour}(2013)}]{Kaltenegger+2013}
{Kaltenegger}, L., \& {Haghighipour}, N. 2013, \apj, 777, 165

\bibitem[{{Kasting} {et~al.}(1993){Kasting}, {Whitmire}, \&
  {Reynolds}}]{Kasting+1993}
{Kasting}, J.~F., {Whitmire}, D.~P., \& {Reynolds}, R.~T. 1993, \icarus, 101,
  108

\bibitem[{{Kodama} {et~al.}(2018){Kodama}, {Nitta}, {Genda}, {Takao}, {O'ishi},
  {Abe-Ouchi}, \& {Abe}}]{Kodama+2018}
{Kodama}, T., {Nitta}, A., {Genda}, H., {et~al.} 2018, Journal of Geophysical
  Research (Planets), 123, 559

\bibitem[{Koll \& Cronin(2018)}]{Koll+2018}
Koll, D. D.~B., \& Cronin, T.~W. 2018, Proceedings of the National Academy of
  Sciences, http://www.pnas.org/content/early/2018/09/24/1809868115.full.pdf

\bibitem[{{Kopparapu} {et~al.}(2017){Kopparapu}, {Wolf}, {Arney}, {Batalha},
  {Haqq-Misra}, {Grimm}, \& {Heng}}]{Kopparapu+2017}
{Kopparapu}, R.~k., {Wolf}, E.~T., {Arney}, G., {et~al.} 2017, \apj, 845, 5

\bibitem[{{Kopparapu} {et~al.}(2016){Kopparapu}, {Wolf}, {Haqq-Misra}, {Yang},
  {Kasting}, {Meadows}, {Terrien}, \& {Mahadevan}}]{Kopparapu+2016}
{Kopparapu}, R.~k., {Wolf}, E.~T., {Haqq-Misra}, J., {et~al.} 2016, \apj, 819,
  84

\bibitem[{{Kopparapu} {et~al.}(2013){Kopparapu}, {Ramirez}, {Kasting}, {Eymet},
  {Robinson}, {Mahadevan}, {Terrien}, {Domagal-Goldman}, {Meadows}, \&
  {Deshpande}}]{Kopparapu+2013}
{Kopparapu}, R.~K., {Ramirez}, R., {Kasting}, J.~F., {et~al.} 2013, \apj, 765,
  131

\bibitem[{{Lacis} \& {Oinas}(1991)}]{Lacis_Oinas_1991}
{Lacis}, A.~A., \& {Oinas}, V. 1991, \jgr, 96, 9027

\bibitem[{{Leconte} {et~al.}(2013){Leconte}, {Forget}, {Charnay}, {Wordsworth},
  {Selsis}, {Millour}, \& {Spiga}}]{Leconte+2013}
{Leconte}, J., {Forget}, F., {Charnay}, B., {et~al.} 2013, \aap, 554, A69

\bibitem[{{Luger} \& {Barnes}(2015)}]{Luger+2015}
{Luger}, R., \& {Barnes}, R. 2015, Astrobiology, 15, 119

\bibitem[{{Manabe} \& {Wetherald}(1967)}]{Manabe_Wetherald_1967}
{Manabe}, S., \& {Wetherald}, R.~T. 1967, Journal of Atmospheric Sciences, 24,
  241

\bibitem[{{Meadows} {et~al.}(2016){Meadows}, {Arney}, {Schwieterman},
  {Lustig-Yaeger}, {Lincowski}, {Robinson}, {Domagal-Goldman}, {Barnes},
  {Fleming}, {Deitrick}, {Luger}, {Driscoll}, {Quinn}, \&
  {Crisp}}]{Meadows_et_al_2016}
{Meadows}, V.~S., {Arney}, G.~N., {Schwieterman}, E.~W., {et~al.} 2016, ArXiv
  e-prints, arXiv:1608.08620

\bibitem[{Murray \& Dermott(1998)}]{Murray1998}
Murray, C.~D., \& Dermott, S.~F. 1998, Solar System Dynamics (Cambridge
  University Press)

\bibitem[{{Nakajima} {et~al.}(1992){Nakajima}, {Hayashi}, \&
  {Abe}}]{Nakajima+1992}
{Nakajima}, S., {Hayashi}, Y.-Y., \& {Abe}, Y. 1992, Journal of Atmospheric
  Sciences, 49, 2256

\bibitem[{North(1975)}]{North1975}
North, G.~R. 1975, Journal of the Atmospheric Sciences, 32, 2033

\bibitem[{{Oshagh} {et~al.}(2017){Oshagh}, {Heller}, \&
  {Dreizler}}]{Oshagh+2017}
{Oshagh}, M., {Heller}, R., \& {Dreizler}, S. 2017, \mnras, 466, 4683

\bibitem[{{Perrin} \& {Hartmann}(1989)}]{Perrin_Hartmann_1989}
{Perrin}, M.~Y., \& {Hartmann}, J.~M. 1989, \jqsrt, 42, 311

\bibitem[{{Pichardo} {et~al.}(2005){Pichardo}, {Sparke}, \&
  {Aguilar}}]{Pichardo+2005}
{Pichardo}, B., {Sparke}, L.~S., \& {Aguilar}, L.~A. 2005, \mnras, 359, 521

\bibitem[{{Pollard}(1983)}]{Pollard1983}
{Pollard}, D. 1983, \jgr, 88, 7705

\bibitem[{{Raghavan} {et~al.}(2010){Raghavan}, {McAlister}, {Henry}, {Latham},
  {Marcy}, {Mason}, {Gies}, {White}, \& {ten Brummelaar}}]{Raghavan+2010}
{Raghavan}, D., {McAlister}, H.~A., {Henry}, T.~J., {et~al.} 2010, \apjs, 190,
  1

\bibitem[{{Ramirez} \& {Kaltenegger}(2014)}]{Ramirez+2014}
{Ramirez}, R.~M., \& {Kaltenegger}, L. 2014, \apjl, 797, L25

\bibitem[{{Rothman} {et~al.}(2013){Rothman}, {Gordon}, {Babikov}, {Barbe},
  {Chris Benner}, {Bernath}, {Birk}, {Bizzocchi}, {Boudon}, {Brown},
  {Campargue}, {Chance}, {Cohen}, {Coudert}, {Devi}, {Drouin}, {Fayt}, {Flaud},
  {Gamache}, {Harrison}, {Hartmann}, {Hill}, {Hodges}, {Jacquemart}, {Jolly},
  {Lamouroux}, {Le Roy}, {Li}, {Long}, {Lyulin}, {Mackie}, {Massie},
  {Mikhailenko}, {M{\"u}ller}, {Naumenko}, {Nikitin}, {Orphal}, {Perevalov},
  {Perrin}, {Polovtseva}, {Richard}, {Smith}, {Starikova}, {Sung}, {Tashkun},
  {Tennyson}, {Toon}, {Tyuterev}, \& {Wagner}}]{Rothman_et_al_2013}
{Rothman}, L.~S., {Gordon}, I.~E., {Babikov}, Y., {et~al.} 2013, \jqsrt, 130, 4

\bibitem[{{Santerne} {et~al.}(2014){Santerne}, {Hebrard}, {Deleuil}, {Havel},
  {Correia}, {Almenara}, {Alonso}, {Arnold}, {Barros}, {Behrend}, {Bernasconi},
  {Boisse}, {Bonomo}, {Bouchy}, {Bruno}, {Damiani}, {Diaz}, {Gravallon},
  {Guillot}, {Labrevoir}, {Montagnier}, {Moutou}, {Rinner}, {Santos}, {Abe},
  {Audejean}, {Bendjoya}, {Gillier}, {Gregorio}, {Martinez}, {Michelet},
  {Montaigut}, {Poncy}, {Rivet}, {Rousseau}, {Roy}, {Suarez}, {Vanhuysse}, \&
  {Verilhac}}]{Santerne+2014}
{Santerne}, A., {Hebrard}, G., {Deleuil}, M., {et~al.} 2014, VizieR Online Data
  Catalog, 357

\bibitem[{{Spiegel} {et~al.}(2008){Spiegel}, {Menou}, \&
  {Scharf}}]{Spiegel+2008}
{Spiegel}, D.~S., {Menou}, K., \& {Scharf}, C.~A. 2008, \apj, 681, 1609

\bibitem[{{Spiegel} {et~al.}(2009){Spiegel}, {Menou}, \& {Scharf}}]{David+2009}
---. 2009, \apj, 691, 596

\bibitem[{{Spiegel} {et~al.}(2010){Spiegel}, {Raymond}, {Dressing}, {Scharf},
  \& {Mitchell}}]{David+2010}
{Spiegel}, D.~S., {Raymond}, S.~N., {Dressing}, C.~D., {Scharf}, C.~A., \&
  {Mitchell}, J.~L. 2010, \apj, 721, 1308

\bibitem[{{Thebault} \& {Haghighipour}(2015)}]{Thebault2015}
{Thebault}, P., \& {Haghighipour}, N. 2015, {Planet Formation in Binaries}, ed.
  S.~{Jin}, N.~{Haghighipour}, \& W.-H. {Ip}, 309--340

\bibitem[{{Tian} \& {Ida}(2015)}]{Tian+2015}
{Tian}, F., \& {Ida}, S. 2015, Nature Geoscience, 8, 177

\bibitem[{{Turbet} {et~al.}(2016){Turbet}, {Leconte}, {Selsis}, {Bolmont},
  {Forget}, {Ribas}, {Raymond}, \& {Anglada-Escud{\'e}}}]{Turbet+2016}
{Turbet}, M., {Leconte}, J., {Selsis}, F., {et~al.} 2016, \aap, 596, A112

\bibitem[{Turbet {et~al.}(2017)Turbet, Bolmont, Leconte, Forget, Selsis, Tobie,
  Caldas, Naar, \& Gillon}]{Turbet+2017}
Turbet, M., Bolmont, E., Leconte, J., {et~al.} 2017, {Astronomy and
  Astrophysics - A\&A}, submitted for publication in Astronomy \& Astrophysics
  2017

\bibitem[{{Way} {et~al.}(2017){Way}, {Aleinov}, {Amundsen}, {Chandler},
  {Clune}, {Del Genio}, {Fujii}, {Kelley}, {Kiang}, {Sohl}, \&
  {Tsigaridis}}]{Way_et_al_2017}
{Way}, M.~J., {Aleinov}, I., {Amundsen}, D.~S., {et~al.} 2017, \apjs, 231, 12

\bibitem[{{Wordsworth} {et~al.}(2010){Wordsworth}, {Forget}, \&
  {Eymet}}]{Wordsworth_et_al_2010}
{Wordsworth}, R., {Forget}, F., \& {Eymet}, V. 2010, \icarus, 210, 992

\bibitem[{{Wordsworth} \& {Pierrehumbert}(2014)}]{Wordsworth+2014}
{Wordsworth}, R., \& {Pierrehumbert}, R. 2014, \apjl, 785, L20

\bibitem[{{Yang} \& {Abbot}(2014)}]{Yang&Abbot2014}
{Yang}, J., \& {Abbot}, D.~S. 2014, \apj, 784, 155

\bibitem[{Yang {et~al.}(2013)Yang, Cowan, \& Abbot}]{Yang+2013}
Yang, J., Cowan, N.~B., \& Abbot, D.~S. 2013, The Astrophysical Journal
  Letters, 771, L45

\bibitem[{{Yokohata} {et~al.}(2002){Yokohata}, {Odaka}, \&
  {Kuramoto}}]{Yokohata+2002}
{Yokohata}, T., {Odaka}, M., \& {Kuramoto}, K. 2002, \icarus, 159, 439

\bibitem[{Young(1980)}]{Young1980}
Young, A.~T. 1980, Appl. Opt., 19, 3427

\end{thebibliography}
\end{document}